\documentclass[preprint,12pt]{elsarticle}

\usepackage{amssymb}
\usepackage{url}

\journal{Finite Fields and Their Applications}

\begin{document}

\begin{frontmatter}



\title{Negacyclic codes over the local ring $\mathbb{Z}_4[v]/\langle v^2+2v\rangle$
of oddly even length and their Gray images}


\author{Yuan Cao, Yonglin Cao$^{\ast}$}

\address{School of Mathematics and  Statistics,
Shandong University of Technology, Zibo, Shandong 255091, China
}
\cortext[cor1]{corresponding author.  \\
E-mail addresses: ylcao@sdut.edu.cn (Yonglin Cao), \ yuancao@sdut.edu.cn (Yuan Cao).}

\begin{abstract}
Let $R=\mathbb{Z}_{4}[v]/\langle v^2+2v\rangle=\mathbb{Z}_{4}+v\mathbb{Z}_{4}$ ($v^2=2v$) and $n$ be an odd positive integer.
Then $R$ is a local non-principal  ideal ring of $16$ elements  and there is a $\mathbb{Z}_{4}$-linear Gray map from $R$ onto $\mathbb{Z}_{4}^2$ which preserves Lee distance and orthogonality.
First, a canonical form decomposition and the structure for any negacyclic code over $R$ of length $2n$ are presented. From this decomposition, a complete classification of all these codes is obtained. Then the cardinality and the dual code
for each of these codes are given, and self-dual negacyclic codes over $R$ of length $2n$ are presented. Moreover, all 
$23\cdot(4^p+5\cdot 2^p+9)^{\frac{2^{p}-2}{p}}$ negacyclic codes over $R$ of length $2M_p$ and all $3\cdot(4^p+5\cdot 2^p+9)^{\frac{2^{p-1}-1}{p}}$ self-dual codes among them are presented precisely, where $M_p=2^p-1$ is a Mersenne prime. Finally, $36$ new and good self-dual $2$-quasi-twisted linear codes over $\mathbb{Z}_4$ with basic parameters $(28,2^{28},
d_L=8,d_E=12)$ and of type $2^{14}4^7$ and basic parameters $(28,2^{28},
d_L=6,d_E=12)$ and of type $2^{16}4^6$ which are Gray images of self-dual negacyclic codes over $R$ of length $14$ are listed.
\end{abstract}


\end{frontmatter}


\section{Introduction}
\noindent
 The catalyst for the study of codes over rings was the discovery of the connection
between the Kerdock and Preparata codes, which are non-linear binary
codes, and linear codes over $\mathbb{Z}_4$ (see [3] and [4]). Soon after this discovery,
codes over many different rings were studied. This led to many new
discoveries and concreted the study of codes over rings as an important
part of the coding theory discipline. Since $\mathbb{Z}_4$ is a chain ring, it was natural to expand the theory to focus on alphabets that are finite commutative chain rings and other special rings (See [1], [2], [5], [7--12],
 [14], [16--22], for examples).

\par
   In 1999, Wood in [23] showed that
for certain reasons finite Frobenius rings are the most general class of
rings that should be used for alphabets of codes. Then self-dual codes over commutative
Frobenius rings were investigated by Dougherty et al. [13]. Especially,
in 2014,
  codes over an extension ring of $\mathbb{Z}_4$ were studied in [24] and
[25], here the
ring  was described as $\mathbb{Z}_4[u]/\langle u^2\rangle=\mathbb{Z}_4+u\mathbb{Z}_4$ ($u^2=0$) which is a local non-principal
ring.

\par
    In this paper, all rings are associative and commutative. Let $A$ be an arbitrary finite ring with identity $1\neq 0$, $A^{\times}$ the multiplicative group of units of
$A$ and $a\in
A$. We denote by $\langle a\rangle_A$, or $\langle a\rangle$ for
simplicity, the ideal of $A$ generated by $a$, i.e. $\langle
a\rangle_A=aA$. For any ideal $I$ of $A$, we will identify the
element $a+I$ of the residue class ring $A/I$ with $a$ (mod $I$) in this paper.

\par
   For any positive integer $N$, let
$A^N=\{(a_0,a_1,\ldots,a_{N-1})\mid a_i\in A, \ i=0,1,\ldots, N-1\}$ which is an $A$-module with componentwise addition and scalar multiplication by elements of $A$. Then an $A$-submodule ${\cal C}$ of $A^N$ is called a \textit{linear code} of length $N$ over $A$.
For any vectors $a=(a_0,a_1,\ldots,a_{N-1}), b=(b_0,b_1,\ldots,b_{N-1})\in A^N$.
The usual \textit{Euclidian inner product} of $a$ and $b$ is defined by
$[a,b]=\sum_{j=0}^{N-1}a_jb_j\in A$. Let ${\cal C}$ be a linear code over $A$ of length $N$. The \textit{dual code}
of ${\cal C}$ is defined by ${\cal C}^{\bot}=\{a\in A^N\mid [a,b]=0, \ \forall
b\in {\cal C}\}$, and ${\cal C}$ is said to be \textit{self-dual} if ${\cal C}={\cal C}^{\bot}$.

\par
  A linear code ${\cal C}$ over $A$ of length $N$ is said to be \textit{negacyclic} if
$$(-a_{N-1},a_0,a_1, \ldots,a_{N-2})\in {\cal C}, \
\forall (a_0,a_1,\ldots,a_{N-1})\in{\cal C}.$$
We will use the natural connection of negacyclic codes
to polynomial rings, where $c=(c_0,c_1,c_2,\ldots$, $c_{N-1})\in A^N$ is viewed as
$c(x)=\sum_{j=0}^{N-1}c_jx^j$ and the negacyclic code ${\cal C}$ is an ideal in the polynomial residue ring
$A[x]/\langle x^N+1\rangle$.

\par
   In this paper, let $n$ be an odd positive integer and denote
$$R=\mathbb{Z}_{4}[v]/\langle v^2+2v\rangle=\{a+bv\mid a,b\in \mathbb{Z}_{4}\}=\mathbb{Z}_{4}+v\mathbb{Z}_{4}
\ (v^2=2v)$$
in which the operations are defined by:
\begin{center}
$\alpha+\beta=(a+b)+v(c+d)$ and $\alpha\beta=ac+(ad+bc+2bd)v$.
\end{center}

\noindent
for any $\alpha=a+bv,\beta=c+dv\in \mathbb{Z}_{4}+v\mathbb{Z}_{4}$ with $a,b,c,d\in \mathbb{Z}_{4}$.
Then $R$ is a local Frobenius non-principal ideal
ring of $16$ elements.

\par
  Linear codes over $R$ were studied in [15]. In the paper, a duality preserving
Gray map was given and used to present MacWilliams identities and self-dual
codes. Some extremal Type II $\mathbb{Z}_{4}$-codes were provided as images of
codes over this ring. $\mathbb{Z}_{4}$-codes that are images of linear codes over $R$ were characterised and some well-known
families of $\mathbb{Z}_{4}$-codes were
proved to be linear over $R$. As in [15] Section 3, we define a map $\varrho: R\rightarrow \mathbb{Z}_4^2$ by
$$\varrho(\alpha)=(a+b,b), \ \forall \alpha=a+bv\in R \ {\rm where} \ a,b\in \mathbb{Z}_4$$
and let $\theta: R^N\rightarrow \mathbb{Z}_4^{2N}$ be such that
$\theta(\alpha_1,\ldots,\alpha_N)=(\varrho(\alpha_1),\ldots,\varrho(\alpha_N))$,
for all $\alpha_1,\ldots,\alpha_N\in R.$
Let $w_L$ denote the Lee weight on $\mathbb{Z}_4$ defined by: 
$$w_L(0)=0, \ w_L(1)=w_L(3)=1 \
{\rm and} \ w_L(2)=2.$$ 
We extend $w_L$ on the ring $R$ in a natural way that
$$w_L(a+bv)=w_L(a+b)+w_L(b), \ \forall a,b\in \mathbb{Z}_4.$$
With this distance and Gray map definition, the following conclusions have been verified by Mart\'{\i}nez-Moro et al. [15].

\vskip 3mm \noindent
  {\bf Lemma 1.1} ([15] Theorem 3.1) \textit{Let $\mathcal{C}$ be a linear code over $R$ of length $N$ and minimum Lee distance $d$. Then $\theta(\mathcal{C})$ is a linear code over $\mathbb{Z}_4$ of length $2N$, $|\theta(\mathcal{C})|=|\mathcal{C}|$ and is of minimum Lee distance $d$}.

\vskip 3mm \noindent
  {\bf Lemma 1.2} ([15] Proposition 3.3) \textit{Let $\mathcal{C}$ be a linear
code over $R$ of length $N$. Then} $\theta(\mathcal{C}^{\bot})=\theta(\mathcal{C})^{\bot}$.
\textit{In particular, if $\mathcal{C}$ is self-dual, then $\theta(\mathcal{C})$ is an
self-dual code over $\mathbb{Z}_4$ of length $2N$ and has
the same Lee weight distribution}.

\vskip 3mm \par
   Moreover, we have the following properties for Negacyclic codes $R$.

\vskip 3mm \noindent
  {\bf Proposition 1.3} \textit{Let $\mathcal{C}$ be a negacyclic code $R$ of length $N$.
Then $\theta(\mathcal{C})$ is a $2$-quasi-twisted code over $\mathbb{Z}_4$ of length $2N$}.

\vskip 3mm \noindent
  {\bf Proof.} Let $\underline{\alpha}=(\alpha_0,\alpha_1,\ldots,\alpha_{N-1})\in \mathcal{C}$,
where $\alpha_i=a_i+b_iv$, $a_i,b_i\in \mathbb{Z}_4$, for all $i=0,1,\ldots,N-1$. Then
$\theta(\underline{\alpha})=(a_0+b_0,b_0,a_1+b_1,b_1,\ldots,a_{N-1}+b_{N-1}, b_{N-1})\in \theta(\mathcal{C})$. As $\mathcal{C}$ is negacyclic, we have $(-\alpha_{N-1},\alpha_0,\alpha_1,\ldots,\alpha_{N-2})$ $\in \mathcal{C}$. This implies
$((-1)(a_{N-1}+b_{N-1}), -b_{N-1},a_0+b_0,b_0,a_1+b_1,b_1,\ldots,a_{N-2}+b_{N-2}, b_{N-2})\in \theta(\mathcal{C})$. So $\theta(\mathcal{C})$ is a $2$-quasi-twisted $\mathbb{Z}_4$-code of length $2N$.
\hfill
$\Box$

\vskip 3mm\par
  Since $n$ is odd, the map $\varphi: R[x]/\langle x^n-1\rangle
\rightarrow R[x]/\langle x^n+1\rangle$ defined by
$$\varphi(a(x))=a(-x)
=\sum_{j=0}^{n-1}a_j(-x)^j \ (\forall a(x)
=\sum_{j=0}^{n-1}a_jx^j\in R[x]/\langle x^n-1\rangle)$$
is an isomorphism
of rings preserving Lee distance. Hence $\mathcal{C}$ is a negacyclic code
over $R$ of $n$ if and only if there is a unique cyclic code $\mathcal{D}$
over $R$ of length $n$ such that $\varphi(\mathcal{D})=\mathcal{C}$. Moreover, $\mathcal{C}$ and $\mathcal{D}$ has the same Lee weight distribution. A complete classification for cyclic codes over $R$ of odd length
and self-dual codes among them had been studied in [5]. In the paper, some good self-dual codes over $\mathbb{Z}_4$ of length
$30$ and extremal binary self-dual codes with parameters $[60,30,12]$ were obtained from self-dual cyclic codes over $R$ of length $15$.
   In this paper,  we study negacyclic codes over $R$ of length
$2n$.

\vskip3mm \par
   The present paper is organized as follows.  In Section 2, we sketch the basic theory of finite rings and linear codes over finite rings needed in this paper.
In Section 3, we decompose the ring $\mathbb{Z}_4[x]/\langle x^{2n}+1\rangle$
into a direct product of finite chain rings of length $4$. In Section 4,
we give a canonical form decomposition for any negacyclic code over $R$ of length $2n$
and present all distinct codes by their generator sets. Using this decomposition, we
give the number of codewords for each of these codes and an enumeration for all these codes. In Section 5, we present the dual code and its self-duality for each negacyclic code over $R$ of length $2n$. In Section 6, we focus our attention on negacyclic code over $R$ of length $2M_p$, where $p$ is a prime and $M_p=2^p-1$ is a Mersenne prime. Especially, we present explicitly all $293687$ ngeacyclic code over $R$ of length $14$ and $339$ self-dual codes among them. Finally, we obtain $36$ new and good self-dual $2$-quasi-twisted codes
over $\mathbb{Z}_4$ of length $28$.


\section{Preliminaries}
\noindent
  In this section, we sketch the basic theory of finite chain rings and linear codes over finite chain rings needed in this paper.

\vskip 3mm \noindent
  {\bf Lemma 2.1} ([10] Proposition 2.1) \textit{Let $\mathcal{K}$ be a finite
  ring with identity. Then the following conditions are equivalent}:

\par
   (i) \textit{$\mathcal{K}$ is a local ring and the maximal ideal $M$ of $\mathcal{K}$ is principal, i.e. $M=\langle \pi\rangle$
for some $\pi\in \mathcal{K}$};

\par
   (ii) \textit{$\mathcal{K}$ is a local principal ideal ring};

\par
   (iii) \textit{$\mathcal{K}$ is a chain ring with all ideals given by: $\langle \pi^i\rangle=\pi^i\mathcal{K}$, $0\leq i\leq s$,
where $s$ is the nilpotency of $\pi$}.

\vskip 3mm \noindent
  {\bf Lemma 2.2} ([10] Proposition 2.2) \textit{Let $\mathcal{K}$ be a finite chain ring,
with maximal ideal $M=\langle\pi\rangle$, and let $s$ be the
nilpotency of $\pi$. Then}
\par
   (i) \textit{For some prime $p$ and positive integer $m$, $|\mathcal{K}/\langle \pi\rangle|=q$ where $q=p^m$,
$|\mathcal{K}|=q^{s}$, and the characteristic of $\mathcal{K}/\langle \pi\rangle$ and
$\mathcal{K}$ are powers of $p$};
\par
  (ii) \textit{For $i=0,1,\ldots,s$, $|\langle\pi^i\rangle|=q^{s-i}$}.

\vskip 3mm\noindent
    {\bf Lemma 2.3} ([16] Lemma 2.4)\textit{Using the notations in Lemma 2.2, let
$\mathcal{T}\subseteq \mathcal{K}$ be a system of representatives for the equivalence classes of $\mathcal{K}$ under
congruence modulo $\pi$. (Equivalently, we can define $\mathcal{T}$ to be a maximal subset of $\mathcal{K}$ with the property that
$t_1-t_2\not\in \langle \pi\rangle$ for all $t_1,t_2\in \mathcal{T}$, $t_1\neq t_2$.) Then}

\vskip 2mm \par
  (i) \textit{Every element $a$ of $\mathcal{K}$ has a unique $\pi$-expansion: $a=\sum_{j=0}^{s-1}t_j\pi^j$,
$t_0,t_1,\ldots,t_{s-1}\in \mathcal{T}$}.

\vskip 2mm \par
  (ii) \textit{$|\mathcal{K}/\langle \pi\rangle|=|\mathcal{T}|$ and $|\langle \pi^i\rangle|=|\mathcal{T}|^{s-i}$ for
$0\leq i\leq s$}.

\vskip 3mm\par
   From now on, let $\mathcal{K}$ be an arbitrary finite chain ring with
$1\neq 0$, $\pi$ be a fixed generator of the maximal ideal of $\mathcal{K}$ with nilpotency index $4$,
and
$F=\mathcal{K}/\langle\pi \rangle$. In this case, $\mathcal{K}$ is called a \textit{finite chain ring of length $4$}. Using the notations of Lemma 2.3, every element $a\in \mathcal{K}$ has a unique \textit{$\pi$-adic expansion}:
$$t_0+\pi t_1+\pi^2t_2+\pi^3t_3, \ t_0,t_1,t_2,t_3\in \mathcal{T}.$$
 Hence $|\mathcal{K}|=|F|^4$. If $a\neq 0$, the \textit{$\pi$-degree} of $a$ is defined as the least index $j\in\{0,1,2,3\}$ for which $t_j\neq 0$ and written for $\|a\|_\pi=j$. If $a=0$
we write $\|a\|_\pi=4$. It is clear that
$a\in \mathcal{K}^{\times}$ if and only if $t_0\neq 0$, i.e. $\|a\|_\pi=0$. Hence $|\mathcal{K}^{\times}|=(|F|-1)|F|^3$. Moreover,
we have $\mathcal{K}/\langle \pi^0\rangle=\{0\}$ and
$\mathcal{K}/\langle \pi^l\rangle=\{\sum_{i=0}^{l-1}\pi^ia_i\mid a_0,\ldots,a_{l-1}\in {\cal T}\}$ with $|\mathcal{K}/\langle \pi^l\rangle|=|F|^l$,
$1\leq l\leq 3$.

\par
   Let $L$ be a positive integer and $\mathcal{K}^L=\{(\alpha_1,\ldots,\alpha_L)\mid \alpha_1,\ldots,\alpha_L\in \mathcal{K}\}$
that is a free $\mathcal{K}$-module under
componentwise addition and scalar multiplication with elements from $\mathcal{K}$.
Then $\mathcal{K}$-submodules of $\mathcal{K}^L$ are linear codes over
$\mathcal{K}$ of length $L$.
   Let $C$ be a linear code over $\mathcal{K}$ of length $L$. By [16]
Definition 3.1, a matrix $G$ is called a \textit{generator matrix} for $C$ if the rows of $G$ span $C$ and none of them can be written as a $\mathcal{K}$-linear combination of the other rows of $G$. Furthermore, a generator matrix $G$ is
said to be \textit{in standard form} if there is a suitable permutation matrix $U$ of size $L\times L$ such that
\begin{equation}
G=\left(\begin{array}{ccccc}
\pi^0I_{k_0} & M_{0,1}     & M_{0,2}       & M_{0,3}       & M_{0,4}\cr
0            & \pi I_{k_1} & \pi M_{1,2}   & \pi M_{1,3}   &  \pi M_{1,4}\cr
0            & 0           & \pi^2 I_{k_2} & \pi^2 M_{2,3} &  \pi^2 M_{2,4}\cr
0            & 0           & 0             & \pi^3 I_{k_3} & \pi^3 M_{3,4} \end{array}\right)U
\end{equation}
where the columns are grouped into blocks of sizes $k_0,k_1,k_2,k_3, k$
with $k_i\geq 0$ and $k=L-(k_0+k_1+k_2+k_3)$. Of course, if $k_i=0$, the matrices $\pi^iI_{k_i}$ and $\pi^iM_{i,j}$ ($i<j$) are suppressed in $G$.
From [16] Proposition 3.2 and Theorem 3.5, we deduce the following.

\vskip 3mm \noindent
   {\bf Lemma 2.4} \textit{Let $C$ be a nonzero linear code of length $L$ over $\mathcal{K}$. Then $C$ has a generator matrix in standard form as in $(1)$. In this case, the number of codewords in $C$ is equal to} $|C|=|F|^{4k_0+3k_1+2k_2+k_3}=|\mathcal{T}|^{4k_0+3k_1+2k_2+k_3}$.

\vskip 3mm\par
   All distinct nontrivial linear codes of length $2$  over
$\mathcal{K}$ has been listed (cf. Cao [6] Lemma 2.2 and Example 2.5). In particular, we have

\vskip 3mm \noindent
   {\bf Theorem 2.5} \textit{Using the notations above, let
$\omega\in \mathcal{K}^{\times}$. Then every linear code $C$ over
$\mathcal{K}$ of length $2$ satisfying the following condition
\begin{equation}
(0, a+\omega \pi^2 b)\in C, \ \forall (a,b)\in C
\end{equation}
has one and only one of the following matrices $G$ as its generator matrix in standard form}:

\vskip 2mm \par
   (I) \textit{$G=(\pi^2(a+b\pi),1)$, where $a,b\in \mathcal{T}$}.

\vskip 2mm \par

\vskip 2mm \par
   (II) \textit{$G=(0,\pi^3)$; $G=(\pi^{3} b,\pi^2)$ where $b\in \mathcal{T}$};
     \textit{$G=(\pi^{3} a,\pi)$  where $a\in \mathcal{T}$}.

\vskip 2mm \par
   (III) \textit{$G=\pi^kI_2$ where $I_2$ is the identity matrix of order $2$, $0\leq k\leq 4$}.

\vskip 2mm \par
   (IV) \textit{$G=\left(\begin{array}{cc}0 & 1\cr
\pi & 0\end{array}\right)$; $G=\left(\begin{array}{cc}\pi^{t-1} z & 1\cr
\pi^t & 0\end{array}\right)$ where $z\in \mathcal{T}$ and $t=2,3$}.

\vskip 2mm \par
   (V) \textit{$G=\left(\begin{array}{cc}\pi^2 z & \pi\cr
\pi^{3} & 0\end{array}\right)$ where $z\in \mathcal{T}$, and $G=\left(\begin{array}{cc}0 & \pi^{t-1}\cr \pi^{t} & 0\end{array}\right)$ where $t=2,3$}.

\vskip 2mm \par
  \textit{Therefore, the number of linear codes over
$\mathcal{K}$ of length $2$ satisfying Condition $(2)$ is equal to $|\mathcal{T}|^2+5|\mathcal{T}|+9$}.

\vskip 3mm \noindent
   {\bf Proof.} See Appendix.
\hfill $\Box$



\section{A direct sum decomposition of the ring $\mathbb{Z}_4[x]/\langle x^{2n}+1\rangle$}
\noindent
From now on, let $n$ be an odd positive integer. In this section, we decompose the ring $\mathbb{Z}_4[x]/\langle x^{2n}+1\rangle$
into a direct product of finite chain rings of length $4$. This direct sum decomposition will be needed in the following
sections.

\par
   It is known that any element $a$ of $\mathbb{Z}_4$ is unique expressed as
$a=a_0+2a_1$ where $a_0,a_1\in \mathbb{F}_2=\{0,1\}$ in which we regard $ \mathbb{F}_2$ as a subset of $\mathbb{Z}_4$. Denote $\overline{a}=a_0\in \mathbb{F}_2$. Then
$^{-}: a\mapsto \overline{a}$ ($\forall a\in \mathbb{Z}_4$) is a ring isomorphism from
$\mathbb{Z}_4$ onto $\mathbb{F}_2$, and $^{-}$ can be extended to a ring isomorphism from
$\mathbb{Z}_4[x]$ onto $\mathbb{F}_2[x]$ by:
$\overline{f}(x)=\overline{f(x)}=\sum_{i=0}^m\overline{b}_ix^i$ for all
$f(x)=\sum_{i=0}^mb_ix^i\in \mathbb{Z}_4[x].$

\par
  A monic polynomial $f(x)\in \mathbb{Z}_4[x]$
is said to be \textit{basic irreducible} if $\overline{f}(x)$ is an irreducible
polynomial in $\mathbb{F}_2[x]$. Then  we have the following conclusions for monic basic irreducible
polynomials in $\mathbb{Z}_4[x]$.

\vskip 3mm \noindent
   {\bf Lemma 3.1} \textit{Let $f(x)$ be a monic basic irreducible polynomial in $\mathbb{Z}_4[x]$ of
degree $m$ and denote $\Gamma=\mathbb{Z}_4[x]/\langle f(x)\rangle$. Then}

\vskip 2mm\par
  (i)  ([22] Theorem 6.1]) \textit{$\Gamma$ is a Galois ring of characteristic $4$ and cardinality $4^{m}$
and $\Gamma=\mathbb{Z}_4[\zeta]$, where $\zeta=x+\langle f(x)\rangle\in \Gamma$ satisfying
$\zeta^{2^{m}-1}=1$ in $\Gamma$}.

\par
   \textit{Denote $\overline{\Gamma}=\mathbb{F}_2[x]/\langle \overline{f}(x)\rangle$ and $\overline{\zeta}=x+\langle \overline{f}(x)\rangle\in \overline{\Gamma}$. Then $\overline{\Gamma}=\mathbb{F}_2[\overline{\zeta}]$ which is a finite field of cardinality $2^m$,
$\overline{f}(x)=\prod_{k=0}^{m-1}(x-\overline{\zeta}^{2^k})$ and that $^{-}$ can be extended to a ring isomorphism from
$\Gamma$ onto $\overline{\Gamma}$ by $\xi\mapsto \overline{\xi}=\sum_{j=0}^{m-1}\overline{a}_j\overline{\zeta}^j$,
for all $\xi=\sum_{j=0}^{m-1}a_j\zeta^j\in \Gamma$ where $a_0,a_1,\ldots,a_{m-1}\in \mathbb{Z}_4$}.

\vskip 2mm\par
  (ii) ([22] Proposition 6.14 or [7] Lemma 2.3(ii)) \textit{$f(x)=\prod_{k=0}^{m-1}(x-\zeta^{2^k})$}.

\vskip 3mm\noindent
   {\bf Theorem 3.2}  \textit{Let $f(x)$ be a monic basic irreducible polynomial in $\mathbb{Z}_4[x]$ of
degree $m$, denote $\mathcal{K}_f=\mathbb{Z}_4[x]/\langle f(-x^2)\rangle=\{\sum_{j=0}^{2m-1}a_jx^j\mid a_0,a_1,\ldots,a_{2m-1}\in\mathbb{Z}_4\}$ in which the arithmetic is done modulo $f(-x^2)$, and set}
$\mathcal{T}_f=\{\sum_{j=0}^{m-1}b_jx^j$ $\mid b_0,b_1,\ldots,b_{m-1}\in\mathbb{F}_2\}\subseteq \mathcal{K}_f$
\textit{in which we regard $\mathbb{F}_2$ as a subset of $\mathbb{Z}_4$. Then}

\vskip 2mm\par
   (i) \textit{There is an invertible element $g(x)\in \mathcal{K}_f$ such that
$f(x)^2=2g(x)$. Hence $\langle f(x)^2\rangle=\langle 2\rangle$ as ideals of $\mathcal{K}_f$}.

\vskip 2mm\par
   (ii) \textit{$\mathcal{K}_f$ is a finite chain ring with maximal ideal $\langle f(x)\rangle$
generated by $f(x)$, the nilpotency
index of $f(x)$ is equal to $4$ and $\mathcal{K}_f/\langle f(x)\rangle$ is a finite field of cardinality $2^{m}$}.

\vskip 2mm\par
   (iii) \textit{Every element $\alpha$ of $\mathcal{K}_f$ has a unique $f(x)$-adic  expansion given by}:
$\alpha=\sum_{j=0}^3b_j(x)f(x)^j, \ b_0(x),b_1(x),b_2(x),b_3(x)\in \mathcal{T}_f.$

\par
   \textit{Moreover, we have $\mathcal{T}_f=\mathcal{K}_f/\langle f(x)\rangle$ as sets and $|\mathcal{T}_f|=2^m$}.

\vskip 3mm\noindent
   {\bf Proof.} (i) By Lemma 3.1(ii), we have
$$f(-x^2)=\prod_{k=0}^{m-1}(-x^2-\zeta^{2^k})=(-1)^{m}\prod_{k=0}^{m-1}(x^2+\zeta^{2^k}).$$
By Lemma 3.1(i), we know that $\zeta^{2^{m}}=\zeta=\zeta^{2^0}$. From this, by Lemma 3.1(ii) and
$(x-\zeta^{2^k})^2=(x^2+\zeta^{2^{k+1}})-2\zeta^{2^k}x$ we deduce that
\begin{eqnarray*}
f(x)^2&=&\prod_{k=0}^{m-1}(x-\zeta^{2^k})^2
=\prod_{k=0}^{m-1}(x^2+\zeta^{2^{k+1}})-2g(x)\\
&=&(-1)^{m}f(-x^2)-2g(x),
\end{eqnarray*}
where $g(x)=x\sum_{k=0}^{m-1}\zeta^{2^k}\prod_{0\leq j\neq k\leq m-1}(x^2+\zeta^{2^{j+1}})\in \mathbb{Z}_4[x]$,
since $g(x)=\frac{1}{2}((-1)^{m}f(-x^2)-f(x)^2)\in \mathbb{Z}[x]$, as a polynomial in
$\mathbb{F}_2[x]$ we have that
$$
\overline{g}(x)=x\sum_{k=0}^{m-1}\overline{\zeta}^{2^k}\prod_{0\leq j\neq k\leq m-1}(x^2+\overline{\zeta}^{2^{j+1}})
=x\sum_{k=0}^{m-1}\overline{\zeta}^{2^k}\prod_{0\leq j\neq k\leq m-1}(x-\overline{\zeta}^{2^j})^2.
$$
This implies $\overline{g}(\overline{\zeta}^{2^k})=\overline{\zeta}^{2^{k+1}}\prod_{0\leq j\neq k\leq m-1}(\overline{\zeta}^{2^k}-\overline{\zeta}^{2^j})^2\neq 0$ for all $k=0,1,2,\ldots$, $m-1$, since
$\overline{\zeta},\overline{\zeta}^2,\ldots,\overline{\zeta}^{2^{m-1}}$ are distinct root of $\overline{f}(x)$ in the finite field $\overline{\Gamma}$ by Lemma 3.1(i). From this and
by
$\overline{f(-x^2)}=\overline{f}(x^2)=(\overline{f}(x))^2=\sum_{k=0}^{m-1}(x-\overline{\zeta}^{2^k})^2$, we deduce that
${\rm gcd}(\overline{f(-x^2)},\overline{g}(x))=1$ in $\mathbb{F}_2[x]$. Therefore, $f(-x^2)$ and $g(x)$ are
coprime in $\mathbb{Z}_4[x]$. Hence there exist $a(x),b(x)\in \mathbb{Z}_4[x]$ such that
$a(x)g(x)+b(x)f(-x^2)=1.$
This implies that $g(x)$ is an invertible element of the residue class ring $\mathcal{K}_f=\mathbb{Z}_4[x]/\langle f(-x^2)\rangle$
and $g(x)^{-1}=a(x)$ (mod $f(-x^2)$). Then by $f(x)^2=(-1)^{d}f(-x^2)-2g(x)$ in $\mathbb{Z}_4[x]$ it follows
that $f(x)^2=2g(x)$ in ${\cal K}_f$. Hence
$\langle 2\rangle=\langle f(x)^2\rangle$
as ideals of ${\cal R}_f$.

\par
   (ii) Let $M=\langle 2,f(x)\rangle$ be the ideal of ${\cal K}_f$ generated by
$2$ and $f(x)$. Then
$${\cal K}_f/M=(\mathbb{F}_2[x]/\langle \overline{f(-x^2)}\rangle)/\langle \overline{f}(x)\rangle
= (\mathbb{F}_2[x]/\langle \overline{f}(x)^2\rangle)/\langle \overline{f}(x)\rangle= \mathbb{F}_2[x]/\langle\overline{f}(x)\rangle$$
up to natural ring isomorphisms, where $\mathbb{F}_2[x]/\langle\overline{f}(x)\rangle$ is a finite field of $2^m$ elements by Lemma 3.1(i). Hence $M$ is a maximal ideal of ${\cal K}_f$.

\par
   By
$f(x)^4=(2g(x))^2=0$, we see that both $2$ and $f(x)$ are nilpotent elements of ${\cal K}_f$. From this one can verify easily
that every element of ${\cal K}_f\setminus M$ is invertible. This implies that ${\cal R}$ is a local ring with $M$ as its unique
maximal ideal. Furthermore, by $\langle 2\rangle=\langle f(x)^2\rangle$ in (i) we conclude that $M=\langle f(x)\rangle$. Hence
${\cal K}_f/\langle f(x)\rangle\cong \mathbb{F}_2[x]/\langle\overline{f}(x)\rangle$.

\par
   As stated above, by Lemma 2.1 we see that ${\cal K}_f$ is a finite chain ring. Let $s$ be the nilpotency
index of $f(x)$. By Lemma 2.2(i) it follows that $|{\cal K}_f|=|\mathbb{F}_2[x]/\langle\overline{f}(x)\rangle|^s=2^{s m}$.
On the other hand, by ${\deg}(f(-x^2))=2m$ it follows that $|{\cal K}_f|=|\mathbb{Z}_4[x]/\langle f(-x^2)\rangle|=4^{2m}=2^{4m}$.
Therefore, $s=4$.

\par
  (iii) By ${\cal K}_f/\langle f(x)\rangle\cong\mathbb{F}_2[x]/\langle\overline{f}(x)\rangle=\{\sum_{j=0}^{m-1}b_jx^j\mid b_0,b_1,\ldots,b_{m-1}\in\mathbb{F}_2\}$,
we see that $\mathbb{F}_2[x]/\langle\overline{f}(x)\rangle={\cal T}_f\subseteq {\cal K}_f$ as sets. Hence ${\cal T}_f$ is a system of representatives for the equivalence classes of ${\cal K}_f$ under
congruence modulo $f(x)$. Then the conclusion follows from Lemma 2.3(i) immediately.
\hfill $\Box$

\vskip 3mm \par
   In the rest of this paper, let
\begin{equation}
x^n-1=f_1(x)f_2(x)\ldots f_r(x),
\end{equation}
where $f_1(x),f_2(x),\ldots, f_r(x)$ are pairwise coprime monic basic irreducible polynomials
in $\mathbb{Z}_4[x]$. We assume ${\rm deg}(f_i(x))=m_i$ and denote
$${\cal K}_i=\mathbb{Z}_4[x]/\langle f_i(-x^2)\rangle, \
  {\cal T}_i=\{\sum_{j=0}^{m_i-1}b_jx^j\mid b_0,b_1,\ldots,b_{m_i-1}\in\mathbb{F}_2\}\subseteq {\cal K}_i,$$
for each integer $i$, $1\leq i\leq r$. Then by Theorem 3.2, we know that

\par
  $\bullet$  There is an invertible element $g_i(x)\in \mathcal{K}_i$ such that
\begin{equation}
f_i(x)^2=2g_i(x) \ {\rm in} \ \mathcal{K}_i
\end{equation}
where $g_i(x)=\frac{1}{2}((-1)^{m_i}f_i(-x^2)-f_i(x)^2)$ as a polynomial in
$\mathbb{Z}[x]$  (mod $f_i(-x^2)$, mod $2$).
Hence $\langle f_i(x)^2\rangle=\langle 2\rangle$ as ideals of $\mathcal{K}_i$.

\par
   $\bullet$ ${\cal K}_i$ is a finite chain ring with maximal ideal $\langle f_i(x)\rangle$, the nilpotency
index of $f_i(x)$ is equal to $4$ and ${\cal K}_i/\langle f_i(x)\rangle$ is a finite field of cardinality $2^{m_i}$.

\par
   $\bullet$ Every element $\alpha$ of ${\cal K}_i$ has a unique $f_i(x)$-adic  expansion:
$\alpha=b_0(x)+b_1(x)f_i(x)+b_2(x)f_i(x)^2+b_3(x)f_i(x)^3$, where $b_j(x)\in {\cal T}_i$ for all $j=0,1,2,3$.
Moreover, we have $\mathcal{T}_i=\mathcal{K}_i/\langle f_i(x)\rangle$ as sets, $|\mathcal{T}_i|=2^{m_i}$
and $|{\cal K}_i|=4^{2m_i}$.

\vskip 2mm \par
  For each $1\leq i\leq r$, denote $F_i(x)=\frac{x^n-1}{f_i(x)}\in \mathbb{Z}_4[x]$.
Since $F_i(x)$ and $f_i(x)$ are coprime, there are polynomials $a_i(x), b_i(x)\in \mathbb{Z}_4[x]$
such that
\begin{equation}
 a_i(x)F_i(x)+b_i(x)f_i(x)=1.
\end{equation}
Substituting $-x^{2}$ for $x$ in (3) and (5), we obtain
$$-(x^{2n}+1)=f_1(-x^{2})f_2(-x^{2})\ldots f_r(-x^{2})$$
and
$
a_i(-x^{2})F_i(-x^{2})+b_i(-x^{2})f_i(-x^{2})=1
$
in the ring $\mathbb{Z}_4[x]$ respectively. In the rest of this paper, we set
\begin{equation}
\varepsilon_i(x)\equiv a_i(-x^{2})F_i(-x^{2})=1-b_i(-x^{2})f_i(-x^{2})
\ ({\rm mod} \ x^{2n}+1).
\end{equation}
Then from classical ring theory, we deduce the following conclusions.

\vskip 3mm \noindent
   {\bf Theorem 3.3} \textit{Denote ${\cal A}=\mathbb{Z}_4[x]/\langle x^{2n}+1\rangle$. We have the following}:

\vskip 2mm \par
   (i) \textit{$\varepsilon_1(x)+\ldots+\varepsilon_r(x)=1$, $\varepsilon_i(x)^2=\varepsilon_i(x)$
and $\varepsilon_i(x)\varepsilon_j(x)=0$ in the ring ${\cal A}$, for all $1\leq i\neq j\leq r$}.

\vskip 2mm \par
   (ii) \textit{${\cal A}={\cal A}_1\oplus\ldots\oplus {\cal A}_r$, where ${\cal A}_i=\varepsilon_i(x){\cal A}$ and its multiplicative
identity is $\varepsilon_i(x)$. Moreover, this decomposition is a direct sum of rings
in that ${\cal A}_i{\cal A}_j=\{0\}$ for all integers $i$ and $j$, $1\leq i\neq j\leq r$}.

\vskip 2mm \par
   (iii) \textit{For each $1\leq i\leq r$, define a mapping $\phi_i: a(x)\mapsto \varepsilon_i(x)a(x)$
$(\forall a(x)\in {\cal K}_i=\mathbb{Z}_4[x]/\langle f_i(-x^{2})\rangle)$. Then $\phi_i$ is a ring isomorphism
from ${\cal K}_i$ onto ${\cal A}_i$. Hence
$|{\cal A}_i|=16^{m_i}$}.

\vskip 2mm \par
   (iv) \textit{Define $\phi: (a_1(x),\ldots,a_r(x))\mapsto \phi_1(a_1(x))+\ldots+\phi_r(a_r(x))$, i.e.
$$\phi(a_1(x),\ldots,a_r(x))=\sum_{i=1}^r\varepsilon_i(x)a_i(x) \ ({\rm mod} \ x^{2n}+1),$$
for all
$a_i(x)\in {\cal K}_i$, $i=1,\ldots,r$. Then $\phi$ is a ring isomorphism from ${\cal K}_1\times\ldots\times{\cal K}_r$ onto
${\cal A}$}.




\section{Structure of negacyclic codes over $R$ of length $2n$}
In this section, we list all distinct negacyclic codes of length $2n$ over the ring $R=\mathbb{Z}_{4}+v\mathbb{Z}_{4}$ ($v^2=2v$),
i.e. all distinct ideals of the ring $R[x]/\langle x^{2n}+1\rangle$. Using the notation of Theorem
3.3, we denote
$$\mathcal{A}=\mathbb{Z}_{4}[x]/\langle x^{2n}+1\rangle=\{\sum_{j=0}^{2n-1}a_jx^j\mid a_j\in \mathbb{Z}_{4}, \ j=0,1,\ldots,2n-1\}$$
in which the arithmetic is done modulo $x^{2n}+1$, and set
$$\mathcal{A}+v\mathcal{A}=\mathcal{A}[v]/\langle v^2+2v\rangle=\{\alpha+\beta v\mid \alpha,\beta\in \mathcal{A}\} \
(v^2=2v)$$
in which the operations are defined by: for any $\alpha_1,\alpha_2,\beta_1, \beta_2\in\mathcal{A}$,

\vskip 2mm\par
  $(\alpha_1+\beta_1 v)+(\alpha_2+\beta_2 v)=(\alpha_1+\alpha_2)+v(\beta_1+\beta_2)$,

\vskip 2mm\par
  $(\alpha_1+\beta_1 v)(\alpha_2+\beta_2 v)=\alpha_1\alpha_2+v(\alpha_1\beta_2+\beta_1\alpha_2+2\beta_1\beta_2)$.

\vskip 2mm\noindent
Then $\mathcal{A}+v\mathcal{A}$ is a finite commutative ring containing $\mathcal{A}$ as its subring.

\par
  Let $\alpha,\beta\in \mathcal{A}$. Then $\alpha$ and $\beta$ can be uniquely expressed as
$\alpha=\sum_{i=0}^{2n-1}a_ix^i$ and $\beta=\sum_{i=0}^{2n-1}b_ix^i$ respectively, where $a_i,b_i\in \mathbb{Z}_{4}$
for all $i=0,1,\ldots,2n-1$. Now, we define a map $\Xi: \mathcal{A}+v\mathcal{A}\rightarrow R[x]/\langle x^{n}-1\rangle$ by
$$\Xi: \alpha+\beta v\mapsto \sum_{i=0}^{2n-1}\xi_ix^i, \
{\rm where} \ \xi_i=a_i+b_iv\in R, \ i=0,1,\ldots,2n-1.$$
Then one can easily verify the following conclusion.

\vskip 3mm \noindent
  {\bf Lemma 4.1} \textit{The map $\Xi$ defined above is an isomorphism of rings from
$\mathcal{A}+v\mathcal{A}$ onto $R[x]/\langle x^{2n}+1\rangle$}.

\vskip 3mm \par
  In the following, we will identify $\mathcal{A}+v\mathcal{A}$ with $R[x]/\langle x^{2n}-1\rangle$
under the ring isomorphism $\Xi$. Therefore, in order to determine all negacyclic codes
over $R$ of length $2n$, we only need to determine all ideals of the ring $\mathcal{A}+v\mathcal{A}$. To do this, we need to
investigate the structure of the ring $\mathcal{A}+v\mathcal{A}$. In the rest
of the paper, for each integer $1\leq i\leq r$ we denote
$$\mathcal{K}_i+v\mathcal{K}_i=\mathcal{K}_i[v]/\langle v^2+2v\rangle=\{\alpha+\beta v\mid \alpha,\beta\in \mathcal{K}_i\}$$
in which the operations are defined by:
$\xi_1+\xi_2=(\alpha_1+\alpha_2)+v(\beta_1+\beta_2)$
and $\xi_1\xi_2=\alpha_1\alpha_2+v(\alpha_1\beta_2+\beta_1\alpha_2+2\beta_1\beta_2)$,
for any $\xi_1=\alpha_1+\beta_1 v$ and $\xi_2=\alpha_2+\beta_2 v$
with $\alpha_1,\alpha_2,\beta_1, \beta_2\in\mathcal{K}_i$.

\vskip 3mm \par
   We give the structure for any negacyclic code over $R$ of length $2n$.

\vskip 3mm \noindent
  {\bf Theorem 4.2} \textit{Using the notations above, we have the following conclusions}.

\vskip 2mm\par
  (i) \textit{Define $\Phi(\xi_1,\ldots,\xi_r)=\sum_{i=1}^r\varepsilon_i(x)\xi_i$ $({\rm mod} \ x^{2n}+1)$
$(\forall \xi_i\in \mathcal{K}_i+v\mathcal{K}_i, \ i=1,2,\ldots,r)$. Then
$\Phi$ is an isomorphism of rings from $(\mathcal{K}_1+v\mathcal{K}_1)\times(\mathcal{K}_2+v\mathcal{K}_2)\times
\ldots\times(\mathcal{K}_r+v\mathcal{K}_r)$ onto $\mathcal{A}+v\mathcal{A}$}.

\vskip 2mm\par
  (ii)  \textit{$\mathcal{C}$ is a negacyclic code
 over $R$ of length $2n$
if and only if for each integer $i$, $1\leq i\leq r$, there is a unique ideal $C_i$ of the ring $\mathcal{K}_i+v\mathcal{K}_i$
such that}
$$\mathcal{C}=\varepsilon_1(x)C_1\oplus \varepsilon_1(x)C_1\oplus\ldots\oplus \varepsilon_r(x)C_r \ ({\rm mod} \ x^{2n}+1)$$
\textit{where $\varepsilon_i(x)C_i=\{\varepsilon_i(x)\alpha+v \varepsilon_i(x)\beta\mid \alpha+\beta v\in C_i, \ \alpha,\beta \in \mathcal{K}_i\}
\subseteq  \mathcal{A}+v\mathcal{A}$
for all $i=1,2,\ldots,r$}. \textit{Hence the number of codewords in $\mathcal{C}$
is $\prod_{i=1}^{r}|C_i|$}.

\vskip 3mm \noindent
   {\bf Proof} (i) Let $\xi=(\xi_1,\xi_2, \ldots, \xi_r)\in (\mathcal{K}_1+v\mathcal{K}_1)\times(\mathcal{K}_1+v\mathcal{K}_1)\times
\ldots\times(\mathcal{K}_r+v\mathcal{K}_r)$
where $\xi_i=\alpha_i+v\beta_i$ and $\alpha_i,\beta_i\in \mathcal{K}_i$ for all $1\leq i\leq r$.
By the definition $\phi$ defined in Theorem 3.3(iv), we have
\begin{eqnarray*}
\Phi(\xi) &=&\sum_{i=1}^r\varepsilon_i(x)\xi_i
 =\sum_{i=1}^r\varepsilon_i(x)\left(\alpha_i+v\beta_i \right)
 =\sum_{i=1}^r\varepsilon_i(x)\alpha_i+v\sum_{i=1}^r\varepsilon_i(x)\beta_i\\
 &=&\phi(\alpha_1,\alpha_2,\ldots,\alpha_r)+v\phi(\beta_1,\beta_2,\ldots,\beta_r).
\end{eqnarray*}
By Theorem 3.3, we know that $\phi$ is a ring isomorphism from  $\mathcal{K}_1\times\mathcal{K}_2\times\ldots\times\mathcal{K}_r$ onto $\mathcal{A}$. Then for any $\eta=(\eta_1,\eta_2,\ldots,\eta_r)$, where $\eta_i=\gamma_i+v\delta_i$ with
$\gamma_i,\delta_i\in \mathcal{K}_i$ for all $i$, by $\Phi(\eta)=\phi(\gamma_1,\gamma_2,\ldots,\gamma_r)
+v\phi(\delta_1,\delta_2,\ldots,\delta_r)$, $v^2=2v$ and direct calculations
one can easily verify that $\Phi(\xi+\eta)=\Phi(\xi)+\Phi(\eta)$ and \begin{eqnarray*}
&&\Phi(\xi\eta)=\Phi(\xi_1\eta_1,\ldots,\xi_r\eta_r)\\
  &=&\Phi\left(\alpha_1\gamma_1+v(\alpha_1\delta_1+\beta_1\gamma_1+2\beta_1\delta_1),\ldots,
  \alpha_r\gamma_r+v(\alpha_r\delta_r+\beta_r\gamma_r+2\beta_r\delta_r)\right)\\
  &=&\phi(\alpha_1\gamma_1,\ldots,\alpha_r\gamma_r)+v\phi(\alpha_1\delta_1+\beta_1\gamma_1+2\beta_1\delta_1,
    \ldots,\alpha_r\delta_r+\beta_r\gamma_r+2\beta_r\delta_r)\\
  &=&\phi(\alpha_1,\ldots,\alpha_r)\phi(\gamma_1,\ldots,\gamma_r)
  +v(\phi(\alpha_1,\ldots,\alpha_r)\phi(\delta_1,\ldots,\delta_r)\\
  &&+\phi(\beta_1,\ldots,\beta_r)\phi(\gamma_1,\ldots,\gamma_r)
    +2\phi(\beta_1,\ldots,\beta_r)\phi(\delta_1,\ldots,\delta_r)) \\
  &=&(\phi(\alpha_1,\ldots,\alpha_r)+v\phi(\beta_1,\ldots,\beta_r))
     (\phi(\gamma_1,\ldots,\gamma_r)+v\phi(\delta_1,\ldots,\delta_r))\\
  &=&\Phi(\xi)\cdot\Phi(\eta).
\end{eqnarray*} Hence $\Phi$ is a ring isomorphism from
$(\mathcal{K}_1+v\mathcal{K}_1)\times\ldots\times(\mathcal{K}_r+v\mathcal{K}_r)$ onto $\mathcal{A}+v\mathcal{A}$.

\par
  (ii) From the properties of ring isomorphisms
and direct product rings,  by (i) we conclude that $\mathcal{C}$ is a negacyclic code over $R$ of length $2n$, i.e. $\mathcal{C}$ is an ideal of $\mathcal{A}+v\mathcal{A}$,
if and only if for each integer $i$, $1\leq i\leq r$, there is a unique ideal $C_i$ of the ring $\mathcal{K}_i+v\mathcal{K}_i$
such that
\begin{eqnarray*}
\mathcal{C}&=&\Phi(C_1\times C_2\times\ldots\times C_r)
  =\{\Phi(\xi_1,\xi_2, \ldots, \xi_r)\mid \xi_i\in C_i, \ 1\leq i\leq r\}\\
 &=&\{\sum_{i=1}^r\varepsilon_i(x)\xi_i\mid \xi_i\in C_i, \ 1\leq i\leq r\}
 =\sum_{i=1}^r\varepsilon_i(x)\{\xi_i\mid \xi_i\in C_i\}.
\end{eqnarray*}
Hence $\mathcal{C}=\bigoplus_{i=1}^r\varepsilon_i(x)C_i$
and $|\mathcal{C}|=|C_1\times C_2\times\ldots\times C_r|=\prod_{i=1}^r|C_i|$.
\hfill $\Box$

\vskip 3mm\par
   Using the notations of Theorem 3.3, ${\cal C}=\bigoplus_{i=1}^r\varepsilon_i(x)C_i$ is called the \textit{canonical form decomposition} of the negacyclic code ${\cal C}$ over $R$ of length $2n$.

\par
   Obviously, $\mathcal{K}_i+v\mathcal{K}_i$ is a free $\mathcal{K}_i$-module with basis $\{1,v\}$.
Let $\mathcal{K}_i^2=\{(\alpha,\beta)\mid \alpha,\beta\in \mathcal{K}_i\}$. Then $\mathcal{K}_i^2$ is a  free $\mathcal{K}_j$-module
of rank $2$ with the componentwise addition and scalar multiplication. Now, define
$$\sigma: \mathcal{K}_i^2\rightarrow \mathcal{K}_i+v\mathcal{K}_i
\ {\rm via} \ (\alpha,\beta)\mapsto \alpha+\beta v \ (\forall \alpha,\beta\in \mathcal{K}_i).$$
Then $\sigma$ is an isomorphism of $\mathcal{K}_i$-modules from
$\mathcal{K}_i^2$ onto $\mathcal{K}_i+v\mathcal{K}_i$. Moreover, we have
the following key conclusion.

\vskip 3mm \noindent
  {\bf Lemma 4.3} (cf. [5] Lemma 3.4) \textit{Let $1\leq i\leq r$. Then $C_i$ is an ideal of the ring $\mathcal{K}_i+v\mathcal{K}_i$
if and only if there is a unique $\mathcal{K}_i$-submodule $S_i$ of $\mathcal{K}_i^2$ satisfying}
\begin{equation}
(0,\alpha+2\beta)\in S_i, \ \forall (\alpha,\beta)\in S_i
\end{equation}
\textit{such that $\sigma(S_i)=C_i$}.

\vskip 3mm \par
   For any ideal $C_i$ of $\mathcal{K}_i+v\mathcal{K}_i$, its
\textit{annihilating ideal} is defined by
${\rm Ann}(C_i)=\{\beta\in \mathcal{K}_i+v\mathcal{K}_i\mid \alpha\beta=0, \ \forall \alpha\in C_i\}$. Now, we list all distinct ideals and their annihilating ideals of the ring $\mathcal{K}_i+v\mathcal{K}_i$ by the following theorem.

\vskip 3mm \noindent
  {\bf Theorem 4.4} \textit{Let $0\leq i\leq r$.
Then all distinct ideals $C_i$ and their annihilating ideals ${\rm Ann}(C_i)$ of the ring $\mathcal{K}_i+v\mathcal{K}_i$ are given by  the following table}.

\begin{center}
\begin{tabular}{lll|l}\hline
 N     &     $C_i$  & $|C_i|$  & ${\rm Ann}(C_i)$  \\ \hline
$2^{2m_i}$ &  $\langle 2(a(x)+b(x)f_i(x))+v\rangle$ & $4^{2m_i}$  & $\langle 2(1+a(x)+b(x)f_i(x))+v\rangle$ \\
$1$ &  $\langle 2vf_i(x)\rangle$ & $2^{m_i}$  & $\langle f_i(x), v\rangle$ \\
$2^{m_i}$ &  $\langle 2(f_i(x) b(x)+v)\rangle$ & $2^{2m_i}$  & $\langle f_i(x)b(x)+v, 2\rangle$ \\
$2^{m_i}$ &  $\langle f_i(x)(2a(x)+v)\rangle$ & $2^{3m_j}$  & $\langle 2(1+a(x))+v, 2f_i(x)\rangle$ \\
$1$ &  $\langle 1\rangle$ & $4^{4m_i}$  & $\langle 0\rangle$ \\
$1$ &  $\langle f_i(x)\rangle$ & $4^{3m_i}$  & $\langle 2f_i(x)\rangle$ \\
$1$ &  $\langle 2\rangle$ & $4^{2m_i}$  & $\langle 2\rangle$ \\
$1$ &  $\langle 2f_i(x)\rangle$ & $4^{m_i}$  & $\langle f_i(x)\rangle$ \\
$1$ &  $\langle 0\rangle$ & $1$  & $\langle 1\rangle$ \\
$1$ &  $\langle f_i(x),v\rangle$ & $2^{7m_i}$  & $\langle 2vf_i(x)\rangle$ \\
$2^{m_i}$ &  $\langle f_i(x)b(x)+v, 2\rangle$ & $2^{6m_j}$  & $\langle 2(f_i(x) b(x)+v)\rangle$ \\
$2^{m_i}$ &  $\langle 2a(x)+v, 2f_i(x)\rangle$ & $2^{5m_i}$  & $\langle f_i(x)(2(1+a(x))+v)\rangle$ \\
$1$ &  $\langle 2, vf_i(x)\rangle$ & $2^{5m_i}$  & $\langle 2f_i(x), 2v\rangle$ \\
$1$ &  $\langle 2f_i(x), 2v\rangle$ & $2^{3m_i}$  & $\langle 2, vf_i(x)\rangle$ \\
$2^{m_i}$ &  $\langle 2b(x)+vf_i(x), 2f_i(x)\rangle$ & $4^{2m_i}$  & $\langle 2b(x)+vf_i(x), 2f_i(x)\rangle$ \\
\hline
\end{tabular}
\end{center}

\noindent
\textit{where $a(x),b(x)\in \mathcal{T}_i$ and $N$ is the number of ideals in the same row. Then the number of ideals in $\mathcal{K}_i+v\mathcal{K}_i$
is equal to $2^{2m_i}+5\cdot 2^{m_i}+9$}.

\vskip 3mm \noindent
   {\bf Proof} Let $C_i$ be an ideal of $\mathcal{K}_i+v\mathcal{K}_j$. By Lemma 4.3
there is a unique $\mathcal{K}_i$-submodule $S$ of $\mathcal{K}_i^2$ satisfying condition
(7) such that $C_i=\sigma(S)$.

\par
   By Equation (4) there is an invertible element $g_i(x)\in \mathcal{K}_j$ such that $f_i(x)^2=2g_i(x)$.
Let $\omega_i(x)=g_i(x)^{-1}\in \mathcal{K}_j$. Then we have $2=\omega_i(x)f_i(x)^2$. From this we deduce
that $S$ satisfying condition (7) if and only if $S$ satisfies the following condition:
$$(0,\alpha+\omega_i(x)f_i(x)^2\beta)\in S_i, \ \forall (\alpha,\beta)\in S.$$
Recall that $\mathcal{K}_i$-submodules $S$ of $\mathcal{K}_i^2$ are called
linear codes over $\mathcal{K}_i$ of length $2$. Since ${\cal K}_i$ is a finite chain ring with maximal ideal $\langle f_i(x)\rangle$, the nilpotency
index of $f_i(x)$ is equal to $4$ and ${\cal T}_i={\cal K}_i/\langle f_i(x)\rangle$
as sets with cardinality $2^{m_i}$, by Theorem 2.5 we conclude that
$S$ has one of the following matrix $G$ as its generator matrix in standard form:

\par
  (I) $G=(f_i(x)^2(a(x)+b(x)f_i(x)),1)$, where $a(x),b(x)\in \mathcal{T}_i$.
Since $g_i(x)\in \mathcal{K}_i^\times$, we have $2g_i(x)=2\overline{g}_i(x)$ where $0\neq\overline{g}_i(x)\equiv g_i(x)$ (mod $2$,
mod $\overline{f}_i(x)$).
Hence $(a(x),b(x))\mapsto (\overline{g}_i(x)a(x),\overline{g}_i(x)b(x))$ (mod $\overline{f}_i(x)$) is a permutation on the set
$\mathcal{T}_i\times \mathcal{T}_i$.
Then by $f_i(x)^2=2g_i(x)$ we have
\begin{eqnarray*}
C_i&=&\langle \sigma(f_i(x)^2(a(x)+b(x)f_i(x)),1)\rangle=\langle f_i(x)^2(a(x)+b(x)f_i(x))+v\rangle\\
  &=&\langle 2g_i(x)(a(x)+b(x)f_i(x))+v\rangle=\langle 2(a^\prime(x)+b^\prime(x)f_i(x))+v\rangle,
\end{eqnarray*}
where $a^\prime(x)=\overline{g}_i(x)a(x), b^\prime(x)=\overline{g}_i(x)b(x)\in \mathcal{T}_i$.
Moreover, by Lemma 2.4 it follows that $|C_i|=|S|=|\mathcal{T}_i|^{4\cdot 1}=(2^{m_i})^4=4^{2m_i}$.
Obviously, the number of ideals is equal to $|\mathcal{T}_i\times \mathcal{T}_i|=|\mathcal{T}_i|^2=2^{2m_i}$ in this case.

\par
  (II) We have one of the following three subcases:

\par
  (II-1) $G=(0,f_i(x)^3)$. Then $C_i=\langle \sigma(0,f_i(x)^3)\rangle
=\langle vf_i(x)^3\rangle=\langle 2vf_i(x)\rangle$ by Equation (4).
Moreover, by Lemma 2.4 we have $|C_i|=|S|=|\mathcal{T}_i|^{1}=2^{m_i}$.

\par
  (II-2) $G=(f_i(x)^{3} b(x),f_i(x)^2)$ where $b(x)\in \mathcal{T}_i$. In this case, we have $C_i=\langle \sigma(f_i(x)^{3} b(x),f_i(x)^2)\rangle
=\langle f_i(x)^{3} b(x)+vf_i(x)^2\rangle=\langle 2(f_i(x) b(x)+v)\rangle$ by Equation (4).
Moreover, by Lemma 2.4 we have $|C_i|=|S|=|\mathcal{T}_i|^{2\cdot 1}=2^{2m_i}$.

\par
  (II-3) $G=(f_i(x)^{3} a(x),f_i(x))$ where $a(x)\in \mathcal{T}_i$.
By $g_i(x)\in \mathcal{K}_i^\times$, an argument similar to (I) shows that the map
$a(x)\mapsto \overline{g}_i(x)a(x)$ (mod $\overline{f}_i(x)$) is a permutation on the set $\mathcal{T}_i$. Then
by Equation (4) we have
\begin{eqnarray*}
C_i&=&\langle \sigma(f_i(x)^{3} a(x),f_i(x)^2)\rangle
=\langle f_i(x)^{3} a(x)+vf_i(x)\rangle\\
&=&\langle 2f_i(x)g_i(x)a(x)+v f_i(x)\rangle=\langle 2f_i(x)a^\prime(x)+v f_i(x)\rangle,
\end{eqnarray*}
where $a^\prime(x)=\overline{g}_i(x)a(x)\in \mathcal{T}_i$  (mod $\overline{f}_i(x)$).
Moreover, by Lemma 2.4 we have $|C_i|=|S|=|\mathcal{T}_i|^{3\cdot 1}=2^{3m_i}$.

\par
  (III) $G=f_i(x)^kI_2$ where $0\leq k\leq 4$.
In this case, we have $C_i=\langle \sigma(f_i(x)^k,0),\sigma(0,f_i(x)^k)\rangle
=\langle f_i(x)^k, vf_i(x)^k\rangle=\langle f_i(x)^k\rangle$.
Moreover, by Lemma 2.4 we have $|C_i|=|S|=|\mathcal{T}_i|^{(4-k)\cdot 2}=2^{2(4-k)m_i}=4^{(4-k)m_i}$.
Precisely, by Equation (4) it follows that $\langle f_i(x)^2\rangle=\langle 2\rangle$
and $\langle f_i(x)^3\rangle=\langle 2f_i(x)\rangle$.

\par
  (IV) We have one of the following three subcases:

\par
  (IV-1) $G=\left(\begin{array}{cc}0 & 1\cr
f_i(x) & 0\end{array}\right)$. Then
$C_i=\langle \sigma(0,1), \sigma(f_i(x),0)\rangle=\langle f_i(x),v\rangle$.
Moreover, by Lemma 2.4 we have $|C_i|=|S|=|\mathcal{T}_i|^{4\cdot 1+3\cdot 1}=2^{7m_i}$.

\par
  (IV-2) $G=\left(\begin{array}{cc}f_i(x)b(x) & 1\cr
f_i(x)^2 & 0\end{array}\right)$ where $b(x)\in \mathcal{T}_i$. In this case,
we have $C_i=\langle \sigma(f_i(x)b(x), 1)$, $\sigma(f_i(x)^2, 0)\rangle
=\langle f_i(x)b(x)+v, f_i(x)^2\rangle=\langle f_i(x)b(x)+v, 2\rangle$
by Equation (4). Then by Lemma 2.4 we have $|C_i|=|S|=|\mathcal{T}_i|^{4\cdot 1+2\cdot 1}=2^{6m_i}$.

\par
  (IV-3) $G=\left(\begin{array}{cc}f_i(x)^2a(x) & 1\cr
f_i(x)^3 & 0\end{array}\right)$ where $a(x)\in \mathcal{T}_i$. By Equation (4) we have
\begin{eqnarray*}
C_i&=&\langle \sigma(f_i(x)^2a(x), 1), \sigma(f_i(x)^3, 0)\rangle
=\langle f_i(x)^2a(x)+v, f_i(x)^3\rangle\\
 &=&\langle 2g_i(x)a(x)+v, 2f_i(x)\rangle=\langle 2a^\prime(x)+v, 2f_i(x)\rangle,
\end{eqnarray*}
where $a^\prime(x)=\overline{g}_i(x)a(x)\in \mathcal{T}_i$ (mod $\overline{f}_i(x)$). Moreover, by Lemma 2.4 we have $|C_i|=|S|=|\mathcal{T}_i|^{4\cdot 1+1\cdot 1}=2^{5m_i}$.

\par
  (V) We have one of the following three subcases:

\par
  (V-1) $G=\left(\begin{array}{cc}0 & f_i(x)\cr f_i(x)^{2} & 0\end{array}\right)$.
In this case,
we have $C_i=\langle \sigma(0, f_i(x))$, $\sigma(f_i(x)^2, 0)\rangle
=\langle vf_i(x), f_i(x)^2\rangle=\langle 2, vf_i(x)\rangle$
by Equation (4). Moreover, by Lemma 2.4 we have $|C_i|=|S|=|\mathcal{T}_i|^{3\cdot 1+2\cdot 1}=2^{5m_i}$.

\par
  (V-2) $G=\left(\begin{array}{cc}0 & f_i(x)^2\cr f_i(x)^{3} & 0\end{array}\right)$.
In this case,
we have $C_i=\langle \sigma(0, f_i(x)^2)$, $\sigma(f_i(x)^3, 0)\rangle
=\langle vf_i(x)^2, f_i(x)^3\rangle=\langle 2f_i(x), 2v\rangle$
by Equation (4). Moreover, by Lemma 2.4 we have $|C_i|=|S|=|\mathcal{T}_i|^{2\cdot 1+1\cdot 1}=2^{3m_i}$.
\par
  (V-3) $G=\left(\begin{array}{cc}f_i(x)^2 b(x) & f_i(x)\cr
f_i(x)^{3} & 0\end{array}\right)$ where $b(x)\in \mathcal{T}_i$.
Then by Equation (4)
we have
\begin{eqnarray*}
C_i&=&\langle \sigma(f_i(x)^2b(x),f_i(x)), \sigma(f_i(x)^3, 0)\rangle
=\langle f_i(x)^2b(x)+vf_i(x), f_i(x)^3\rangle\\
&=&\langle 2g_i(x)b(x)+vf_i(x), 2f_i(x)\rangle=\langle 2b^\prime(x)+vf_i(x), 2f_i(x)\rangle,
\end{eqnarray*}
where $b^\prime(x)=\overline{g}_i(x)b(x)\in \mathcal{T}_i$ (mod $\overline{f}_i(x)$). Moreover, by Lemma 2.4 we have $|C_i|=|S|=|\mathcal{T}_i|^{3\cdot 1+1\cdot 1}=2^{4m_i}$.
  By Theorem 2.5, we see that the number of ideals in
$\mathcal{K}_i+v\mathcal{K}_i$ is equal to $|\mathcal{T}_i|^2+5|\mathcal{T}_i|+9=2^{2m_i}+5\cdot 2^{m_i}+9$.

\par
  Since $\mathcal{K}_i+v\mathcal{K}_i$ is a local ring with maximal ideal
$\langle f_i(x), v\rangle$, one can deduce the conclusions for annihilating ideals
from $v^2=2v$, $\omega_i(x)f_i(x)^2=2$ and a direct computation.
\hfill $\Box$

\vskip 3mm \par
   Finally, from Theorems 4.2 and 4.4 we deduce the following corollary.

\vskip 3mm \noindent
  {\bf Corollary 4.5} \textit{Every negacyclic code over $R$ of length $2n$ can be constructed by the following two steps}:

\par
  (i) \textit{For each $i=1,\ldots,r$, choose an ideal $C_i$ of
${\cal K}_i+v{\cal K}_i$ listed in Theorem 4.4}.

\par
  (ii) \textit{Set ${\cal C}=\bigoplus_{i=1}^r\varepsilon_i(x)C_i$} (mod $x^{2n}+1$).

\noindent
\textit{The number of codewords in ${\cal C}$ is
equal to $|{\cal C}|=\prod_{i=1}^r|C_i|$}.

\par
   \textit{Moreover, the number of negacyclic
codes over $R$ of length $2n$ is equal to}:
$\prod_{i=1}^r\left(2^{2m_i}+5\cdot 2^{m_i}+9\right).$



\section{Dual codes of negacyclic codes over $R$ of length $2n$}
  In this section, we give the dual code of each negacyclic code over $R=\mathbb{Z}_{4}+v\mathbb{Z}_{4}$
of length $2n$ and investigate the self-duality of these codes.

\par
  \par
   Let $\alpha=(\alpha_0,\alpha_1,\ldots,\alpha_{2n-1}), \beta=(\beta_0,\beta_1,\ldots,\beta_{2n-1})\in R^{2n}$,
where $\alpha_j,\beta_j\in R$ for all $j=0,1\ldots,2n-1$. As usual, we will identify
the vector $\alpha$
with $\alpha(x)=\sum_{j=0}^{2n-1}\alpha_jx^j\in R[x]/\langle x^{2n}+1\rangle$ in this paper.
Then we define
$$\mu(\alpha(x))=\alpha(x^{-1})=\alpha_0-\sum_{j=1}^{2n-1}\alpha_jx^{2n-j}\in R[x]/\langle x^{2n}+1\rangle.$$
It is clear that $\mu$ is a ring automorphism of $R[x]/\langle x^{2n}+1\rangle$ satisfying
$\mu^{-1}=\mu$. Now, by a direct calculation we get the following lemma.

\vskip 3mm \noindent
  {\bf Lemma 5.1} \textit{Let $\alpha,\beta\in R^{2n}$.
Then $[\alpha,\beta]=\sum_{j=0}^{2n-1}\alpha_j\beta_j=0$  if $\alpha(x)\mu(\beta(x))=0$
in $R[x]/\langle x^{2n}+1\rangle$}.

\vskip 3mm \par
   Using the notations of Section 4, we have $R[x]/\langle x^{2n}+1\rangle=
{\cal A}+v{\cal A}$ where ${\cal A}=\mathbb{Z}_4[x]/\langle x^{2n}+1\rangle$ and $v^2=2v$. It is obvious that the restriction of $\mu$
to ${\cal A}$ is a ring automorphism of ${\cal A}$. We still denote this automorphism by $\mu$, i.e.
$\mu(a(x))=a(x^{-1})$ for any $a(x)\in {\cal A}$. Let $1\leq i\leq r$. By Equation (6) in Section 3, we have
\begin{equation}
\mu(\varepsilon_i(x))=a_i(-x^{-2})F_i(-x^{-2})=1-b_i(-x^{-2})f_i(-x^{-2}) \
{\rm in} \ {\cal A}
\end{equation}
For any polynomial $f(x)=\sum_{j=0}^mc_jx^j\in \mathbb{Z}_4[x]$ of degree $m\geq 1$, recall that
the \textit{reciprocal polynomial} of $f(x)$ is defined as $\widetilde{f}(x)=x^mf(\frac{1}{x})=\sum_{j=0}^mc_jx^{m-j}$, and
 $f(x)$ is said to be \textit{self-reciprocal} if $\widetilde{f}(x)=f(x)$ or $-f(x)$. Then by Equation (3) in Section 3, we have
$$x^n-1=-\widetilde{(x^n-1)}=-\widetilde{f}_1(x)\widetilde{f}_2(x)\ldots \widetilde{f}_r(x).$$
Since $f_1(x),f_2(x),\ldots,f_r(x)$ are pairwise coprime monic basic irreducible polynomials in $\mathbb{Z}_4[x]$,
$\widetilde{f}_1(x),\widetilde{f}_2(x),\ldots, \widetilde{f}_r(x)$  are pairwise coprime basic irreducible polynomials in $\mathbb{Z}_4[x]$ as well. Hence for each integer $i$, $1\leq i\leq r$,
there is a unique integer $i^{\prime}$, $1\leq i^{\prime}\leq r$, such that $\widetilde{f}_i(x)=\delta_if_{i^{\prime}}(x)$
where $\delta_i\in \{1,-1\}$.
  We assume that ${\rm deg}(f_i(x))=d_i$ and $f_i(x)=\sum_{j=0}^{m_i}c_jx^j$ where $c_j\in \mathbb{Z}_4$. Then
\begin{eqnarray*}
x^{2m_i}f_i(-x^{-2})&=&(x^{2})^{m_i}\sum_{j=0}^{m_i}(-1)^jc_j(x^{2})^{-j}
=(-1)^{m_i}\sum_{j=0}^{m_i}c_j(-x^{2})^{m_i-j}\\
&=&(-1)^{m_i}\widetilde{f}_i(-x^{2})=(-1)^{m_i}\delta_if_{i^{\prime}}(-x^{2}).
\end{eqnarray*}
  From this, by Equation (6) and $x^{2n}=-1$ in the ring ${\cal A}$, we deduce that
\begin{eqnarray*}
\mu(\varepsilon_i(x))&=&1+x^{2n-2({\rm deg}(b_i(x))+m_i)}(x^{2{\rm deg}(b_i(x))}b_i(-x^{-2}))(x^{2m_i}f_i(-x^{-2}))\\
  &=&1+(-1)^{{\rm deg}(b_i(x))+m_i}x^{2n-2({\rm deg}(b_i(x))+m_i)}\widetilde{b}_i(-x^{2})\widetilde{f}_i(-x^{2})\\
  &=&1-h_i(x)f_{i^{\prime}}(-x^{2})
\end{eqnarray*}
where $h_i(x)=(-1)^{{\rm deg}(b_i(x))+m_i+1}\delta_ix^{2n-2({\rm deg}(b_i(x))+m_i)}\widetilde{b}_i(-x^{2})\in {\cal A}$. Similarly, by (6) it
follows that $\mu(\varepsilon_i(x))=g_i(x)F_{i^{\prime}}(-x^{2})$ for some $g_i(x)\in {\cal A}$. Then from these
and by (6), we deduce that $\mu(\varepsilon_i(x))=\varepsilon_{i^{\prime}}(x)$.

\par
   As stated above, we see that
for each $1\leq i\leq r$ there is a unique integer $i^{\prime}$, $1\leq i^{\prime}\leq r$, such that $\mu(\varepsilon_i(x))=
\varepsilon_{i^{\prime}}(x)$. We still use $\mu$ to denote this map $i\mapsto i^{\prime}$. Then $\mu(\varepsilon_i(x))=\varepsilon_{\mu(i)}(x)$.

\par
   Whether $\mu$ denotes the automorphism of ${\cal A}$ or this map on the set $\{1,2$, $\ldots,r\}$ can be determined by the context.
The next lemma shows the compatibility of the two uses of $\mu$.

\vskip 3mm \noindent
  {\bf Lemma 5.2} \textit{Using the notations above, we have the following conclusions}.

\vskip 2mm \par
   (i) \textit{$\mu$ is a permutation on $\{1,\ldots,r\}$ satisfying $\mu^{-1}=\mu$}.

\vskip 2mm \par
   (ii) \textit{After a rearrangement of $f_1(x),\ldots,f_r(x)$ there are integers $\lambda$ and $\epsilon$ such that
$\mu(i)=i$ for all $i=1,\ldots,\lambda$ and $\mu(\lambda+j)=\lambda+\epsilon+j$ for all $j=1,\ldots,\epsilon$, where $\lambda\geq 1,
\epsilon\geq 0$ and $\lambda+2\epsilon=r$}.

\vskip 2mm\par
   (iii) \textit{For each integer $i$, $1\leq i\leq r$, there is a unique element $\delta_i$ of
$\{1,-1\}$ such that $\widetilde{f}_i(x)=\delta_i f_{\mu(i)}(x)$}.

\vskip 2mm \par
   (iv) \textit{For any integer $i$, $1\leq i\leq r$, $\mu(\varepsilon_i(x))=\varepsilon_{\mu(i)}(x)$
and $\mu({\cal A}_{i})={\cal A}_{\mu(i)}$ in the ring ${\cal A}$}.

\vskip 2mm \par
   (v) \textit{For any $c(x)\in {\cal K}_i=\mathbb{Z}_4[x]/\langle f_i(-x^2)\rangle$, define
$$\mu_i(c(x))=c(x^{-1})=c(-x^{2n-1}) \ ({\rm mod} \ f_{\mu(i)}(-x^2)).$$
Then $\mu_i=\phi_{\mu(i)}^{-1}\mu\phi_i$ which is a ring isomorphism from
${\cal K}_i$ onto ${\cal K}_{\mu(i)}$}.

\vskip 2mm \par
   (vi) \textit{The ring isomorphism $\mu_i: {\cal K}_i\rightarrow {\cal K}_{\mu(i)}$ induces a ring isomorphism from ${\cal K}_{i}+v{\cal K}_{i}$
 onto ${\cal K}_{\mu(i)}+v{\cal K}_{\mu(i)}$ $(v^2=2v)$ defined by
$$\alpha+v\beta\mapsto \mu_i(\alpha)+v\mu_i(\beta), \
\forall \alpha,\beta\in{\cal K}_i.$$
We still denote this isomorphism by $\mu_i$. Then $\mu_i^{-1}=\mu_{\mu(i)}$}.

\vskip 3mm \noindent
  {\bf Proof.} (i)--(iii) follow from the definition of the map $\mu$.

\par
   (iv) By $\mu(\varepsilon_i(x))=\varepsilon_{\mu(i)}(x)$ and ${\cal A}_i=\varepsilon_i(x){\cal A}$, it follows that
$\mu({\cal A}_i)=\mu(\varepsilon_i(x))\mu({\cal A})=\varepsilon_{\mu(i)}(x){\cal A}={\cal A}_{\mu(i)}$.

\par
   (v) Since $x^{2n}=-1$ in the ring $\mathcal{A}$, i.e. $x^{2n}\equiv -1$ (mod $x^{2n}+1$),
and $f_i(-x^2)$ is a factor of $x^{2n}+1$, it follows that $x^{2n}\equiv -1$ (mod $f_i(-x^2)$).
This implies $x^{-1}=-x^{2n-1}$ in the ring $\mathcal{K}_i$ for all $i$, $1\leq i\leq r$. Now,
   let $c(x)\in {\cal K}_i$. By Theorem 3.3(iii) and $\varepsilon_{\mu(i)}(x)=\mu(\varepsilon_i(x))=1-h_i(x)f_{i^{\prime}}(-x^2)
=1-h_i(x)f_{\mu(i)}(-x^2)$, we have
\begin{eqnarray*}
(\phi_{\mu(i)}^{-1}\mu\phi_i)(c(x))&=&(\phi_{\mu(i)}^{-1}\mu)(\varepsilon_i(x)c(x))
=\phi_{\mu(i)}^{-1}(\mu(\varepsilon_i(x))c(x^{-1})) \\
&=&(1-h_i(x)f_{\mu(i)}(-x^2))c(x^{-1})  \ ({\rm mod} \ f_{\mu(i)}(-x^2)) \\
&\equiv& c(x^{-1}) \ ({\rm mod} \ f_{\mu(i)}(-x^2)).
\end{eqnarray*}
This implies $\mu_i(c(x))=(\phi_{\mu(i)}^{-1}\mu\phi_i)(c(x))$ for all $c(x)\in {\cal R}_i$. Hence
$\mu_i=\phi_{\mu(i)}^{-1}\mu\phi_i$. Since $\mu$ is a ring automorphism of $\mathcal{A}$,
by Theorem 3.3(iii) we conclude that $\mu_i$  is a ring isomorphism from
${\cal K}_i$ onto ${\cal K}_{\mu(i)}$.

\par
  (vi) Obviously, $\mu_i$ can be extended to a ring isomorphism from ${\cal K}_i[v]$ onto ${\cal K}_{\mu(i)}[v]$ in
the natural way that
$\mu_i:\sum_j\alpha_jv^j\mapsto \sum_j\mu_i(\alpha_j)v^j$, 
$\forall \alpha_j\in {\cal K}_i.$
Therefore,
$\alpha+v\beta\mapsto \mu_i(\alpha)+v\mu_i(\beta)$ $(\forall \alpha,\beta\in{\cal K}_i)$ is a
ring isomorphism from ${\cal K}_{i}[v]/\langle v^2+2v\rangle={\cal K}_{i}+v{\cal K}_{i}$
 onto ${\cal K}_{\mu(i)}[v]/\langle v^2+2v\rangle={\cal K}_{\mu(i)}+v{\cal K}_{\mu(i)}$.
\hfill $\Box$

\vskip 3mm\par
   By Theorem 4.2(i), each element $\xi\in \mathcal{A}+v\mathcal{A}$ can be uniquely
expressed as: $\xi=\sum_{i=1}^r\phi_i(\xi_i)=\sum_{i=1}^r\varepsilon_i(x)\xi_i$,
where $\xi_i\in \mathcal{K}_i+v\mathcal{K}_i$ for all $i=1,\ldots,r$.

\vskip 3mm \noindent
  {\bf Lemma 5.3} \textit{Let $\xi=\sum_{i=1}^r\varepsilon_i(x)\xi_i, \eta=\sum_{i=1}^r\varepsilon_i(x)\eta_i\in {\cal A}+v{\cal A}$,
where $\xi_i, \eta_i\in{\cal K}_i+v{\cal K}_i$. Then
$\xi\cdot\mu(\eta)=\sum_{i=1}^r\varepsilon_i(x)(\xi_i\cdot\mu_i^{-1}(\eta_{\mu(i)}))$}.

\vskip 3mm \noindent
  {\bf Proof.} By Lemma 5.2(v) we have $\mu_i^{-1}(\eta_{\mu(i)})\in \mu_i^{-1}({\cal K}_{\mu(i)}+v{\cal K}_{\mu(i)})={\cal K}_i+v{\cal K}_i$.
Hence $\xi_i\cdot\mu_i^{-1}(\eta_{\mu(i)})\in {\cal K}_i+v{\cal K}_i$ for all $i$. If
$j\neq \mu(i)$, then $i\neq\mu(j)$, which implies $\varepsilon_i(x)\varepsilon_{\mu(j)}(x)=0$ by Theorem 3.3(i). Therefore,
\begin{eqnarray*}
\xi\cdot\mu(\eta)&=&\sum_{i,j=1}^r\varepsilon_i(x)\xi_i\cdot\mu(\varepsilon_j(x)\eta_j)
=\sum_{i,j=1}^r\varepsilon_i(x)\xi_i\cdot\mu(\varepsilon_j(x))\mu_j(\eta_j)\\
&=&\sum_{i,j=1}^r\varepsilon_i(x)\xi_i\cdot\varepsilon_{\mu(j)}(x)\mu_j(\eta_j)
=\sum_{i=1}^r\varepsilon_i(x)\xi_i\cdot\varepsilon_{i}(x)\mu_{\mu(i)}(\eta_{\mu(i)}).
\end{eqnarray*}
This implies $\xi\cdot\mu(\eta)=\sum_{i=1}^r\varepsilon_i(x)(\xi_i\cdot\mu_i^{-1}(\eta_{\mu(i)}))$ by
$\mu_{\mu(i)}=\mu_i^{-1}$.
\hfill $\Box$

\vskip 3mm \noindent
  {\bf Lemma 5.4} \textit{Let $1\leq i\leq r$. Then}
$\mu_i(f_i(x))=-\delta_ix^{2n-m_i}f_{\mu(i)}(x).$

\vskip 3mm \noindent
  {\bf Proof.} Since $f_i(-x^2)|(x^{2n}+1)$ in $\mathbb{Z}_4[x]$, we have $x^{2n}=-1$ in ${\cal K}_i$.
By the definition of $\mu_i$ and $\widetilde{f}_i(x)=\delta_if_{\mu(i)}(x)$ in Lemma 5.2 (v) and (iii), we have that
$\mu_i(f_i(x))=f_i(x^{-1})=-x^{2n-m_i}(x^{m_i}f_i(x^{-1}))=-x^{2n-m_i}\widetilde{f}_i(x)
=-\delta_i x^{2n-m_i}f_{\mu(i)}(x)$.
\hfill $\Box$

\vskip 3mm \par
    Now, we can give the dual code of each negacyclic code over the ring $R=\mathbb{Z}_4+v\mathbb{Z}_4$
of length $2n$ by the following theorem, where $h(x^{-1})=h(-x^{2n-1})$ (mod $f_{\mu(i)}(-x^2)$) for any $h(x)\in \mathcal{K}_i$.

\vskip 3mm \noindent
   {\bf Theorem 5.5} \textit{Let ${\cal C}$ be a negacyclic code
over $R$ of length $2n$ with
${\cal C}=\oplus_{i=1}^r\varepsilon_i(x)C_i$, where $C_i$ is an ideal of ${\cal K}_i+v{\cal K}_i$. Then
the dual code ${\cal C}^{\bot}$ is also a negacyclic code
over $R$ of length $2n$. Precisely, we have}
$${\cal C}^{\bot}=\bigoplus_{j=1}^r\varepsilon_j(x)D_j,$$
\textit{where $D_j$ is an ideal of ${\cal K}_j+v{\cal K}_j$ determined by the following table}:
\begin{center}
\begin{tabular}{ll}\hline
  $C_i$ (mod $f_i(-x^2)$) &  $D_{\mu(i)}$ (mod $f_{\mu(i)}(-x^2)$) \\ \hline
 $\langle 2(a(x)+b(x)f_i(x))+v\rangle$ &  $\langle 2(1+a(x^{-1})-\delta_ix^{2n-m_i}b(x^{-1})f_{\mu(i)}(x))
 +v\rangle$  \\
  $\langle 2vf_i(x)\rangle$ & $\langle f_{\mu(i)}(x),v\rangle$ \\
  $\langle 2(f_i(x) b(x)+v)\rangle$ & $\langle -f_{\mu(i)}(x)\delta_ix^{2n-m_i}b(x^{-1})+v, 2\rangle$ \\
 $\langle f_i(x)(2a(x)+v)\rangle$   & $\langle 2(1+a(x^{-1}))+v, 2f_{\mu(i)}(x)\rangle$ \\
  $\langle 1\rangle$   & $\langle 0\rangle$ \\
  $\langle f_i(x)\rangle$  & $\langle 2f_{\mu(i)}(x)\rangle$ \\
  $\langle 2 \rangle$   & $\langle 2\rangle$ \\
 $\langle 2f_i(x)\rangle$   & $\langle f_{\mu(i)}(x)\rangle$ \\
  $\langle 0\rangle$   & $\langle 1\rangle$ \\
 $\langle f_i(x),v\rangle$   & $\langle 2vf_{\mu(i)}(x)\rangle$ \\
 $\langle f_i(x)b(x)+v, 2\rangle$  & $\langle 2(-f_{\mu(i)}(x)\delta_ix^{2n-m_i} b(x^{-1})+v)\rangle$ \\
 $\langle 2a(x)+v, 2f_i(x)\rangle$ & $\langle f_{\mu(i)}(x)(2(1+a(x^{-1}))+v)\rangle$ \\
 $\langle 2, vf_i(x)\rangle$  & $\langle 2f_{\mu(i)}(x), 2v\rangle$   \\
 $\langle 2f_i(x), 2v\rangle$   & $\langle 2, vf_{\mu(i)}(x)\rangle$ \\
 $\langle 2b(x)+vf_i(x), 2f_i(x)\rangle$ & $\langle 2b(x^{-1})-vf_{\mu(i)}(x)\delta_ix^{2n-m_i}, 2f_{\mu(i)}(x)\rangle$ \\
 \hline
\end{tabular}
\end{center}

\noindent
\textit{where $a(x),b(x)\in \mathcal{T}_i$}.

\vskip 3mm \noindent
  {\bf Proof.} Let $1\leq i\leq r$. By Theorem 4.4 we see that
$C_i\cdot{\rm Ann}(C_i)=\{0\}$ and $|C_i||{\rm Ann}(C_i)|=4^{4m_i}$.
Since $\mu_i$ is a ring isomorphism from $\mathcal{K}_i+v\mathcal{K}_i$ onto
$\mathcal{K}_{\mu(i)}+v\mathcal{K}_{\mu(i)}$, $\mu_i({\rm Ann}(C_i))$ is an ideal
of $\mathcal{K}_{\mu(i)}+v\mathcal{K}_{\mu(i)}$. Denote $D_{\mu(i)}=\mu_i({\rm Ann}(C_i))$ and set $\mathcal{D}=\sum_{i=1}^r\varepsilon_{\mu(i)}(x)D_{\mu(i)}$.
Then $\mathcal{D}=\bigoplus_{j=1}^r\varepsilon_j(x)D_j$ by Lemma 5.2(i) and Theorem 3.3(i). From
this and by Theorem 4.4(ii), we deduce that $\mathcal{D}$ is a negacyclic code over $R$ of length $2n$.

\par
  As $\mu_i^{-1}(D_{\mu(i)})={\rm Ann}(C_i)$ for all $i$, by Lemma 5.3
it follows that
$${\cal C}\cdot \mu({\cal D})=\sum_{i=1}^r\varepsilon_i(x)\left(C_i\cdot\mu_i^{-1}(D_{\mu(i)})\right)
=\sum_{i=1}^r\varepsilon_i(x)\left(C_i\cdot {\rm Ann}(C_i)\right)=\{0\}.$$
This implies ${\cal D}\subseteq {\cal C}^{\bot}$ by Lemma 5.1.
  On the other hand, we have
\begin{eqnarray*}
|{\cal C}||{\cal D}|&=&(\prod_{i=1}^r|C_i|)(\prod_{i=1}^r|D_{\mu(i)}|)
=\prod_{i=1}^r|C_i||{\rm Ann}(C_i)|=4^{4\sum_{i=1}^rm_i}=4^{4n}\\
&=&|(\mathbb{Z}_4+v\mathbb{Z}_4)[x]/\langle x^{2n}+1\rangle|=|(\mathbb{Z}_4+v\mathbb{Z}_4)^{2n}|
\end{eqnarray*}
by Theorem 4.2(ii). Since
$\mathbb{Z}_4+v\mathbb{Z}_4$ is a Frobenius ring, from the theory of linear codes over Frobenius rings (see [13])
we deduce that ${\cal C}^{\bot}={\cal D}$.

\par
  Finally, for any integer $i$, $1\leq i\leq r$, the expression for each ideal
$$D_{\mu(i)}=\mu_i({\rm Ann}(C_i))
=\{\mu_i(h(x)) \mid h(x)\in {\rm Ann}(C_i)\} \ ({\rm mod} \ f_{\mu(i)}(-x^2))$$
follows from Theorem 4.4 and Lemma 5.4 immediately.
\hfill $\Box$

\vskip 3mm \par
   As the end of this section, we list all distinct self-dual negacyclic
codes over $R$ by Theorems 4.2 and 5.5.

\vskip 3mm \noindent
  {\bf Theorem 5.6} \textit{Using the notations in Theorem 5.5 and Lemma 5.2$({\rm ii})$, denote
\begin{eqnarray*}
\mathcal{W}_i^{(1)}&=&\{(a(x),b(x))\mid a(x)+a(x^{-1})+1\equiv b(x)+x^{2n-m_i}b(x^{-1})\equiv 0 \\
 && ({\rm mod} \ 2, {\rm mod} \ \overline{f}_i(x)), \ a(x),b(x)\in \mathbb{F}_2[x]/\langle \overline{f}_i(x)\rangle\}
\end{eqnarray*}
\textit{and}
$$\mathcal{W}_i^{(2)}=\{b(x)\in \mathbb{F}_2[x]/\langle \overline{f}_i(x)\rangle\mid b(x)+x^{m_i}b(x^{-1})\equiv 0 \ ({\rm mod} \ 2, {\rm mod} \ \overline{f}_i(x))\}$$
for any $1\leq i\leq \lambda$.
Let ${\cal C}$ be a negacyclic code
over $R$ of length $2n$ with
${\cal C}=\bigoplus_{i=1}^r\varepsilon_i(x)C_i$, where $C_i$ is an ideal of ${\cal K}_i+v{\cal K}_i$. Then
${\cal C}$ is self-dual if and only if $C_i$ satisfies the following conditions}:

\vskip 2mm \par
  (i) \textit{If $i=\lambda+j$ where $1\leq j\leq \epsilon$, $(C_i,C_{i+\epsilon})$ is given by the following table}:
\begin{center}
\begin{tabular}{ll}\hline
  $C_i$ (mod $f_i(-x^2)$) &  $C_{i+\epsilon}$ (mod $f_{i+\epsilon}(-x^2)$) \\ \hline
 $\langle 2(a(x)+b(x)f_i(x))+v\rangle$ &  $\langle 2(1+a(x^{-1})-\delta_ix^{2n-m_i}b(x^{-1})f_{i+\epsilon}(x))
 +v\rangle$  \\
  $\langle 2vf_i(x)\rangle$ & $\langle f_{i+\epsilon}(x),v\rangle$ \\
  $\langle 2(f_i(x) b(x)+v)\rangle$ & $\langle -f_{i+\epsilon}(x)\delta_ix^{2n-m_i}b(x^{-1})+v, 2\rangle$ \\
 $\langle f_i(x)(2a(x)+v)\rangle$   & $\langle 2(1+a(x^{-1}))+v, 2f_{i+\epsilon}(x)\rangle$ \\
  $\langle 1\rangle$   & $\langle 0\rangle$ \\
  $\langle f_i(x)\rangle$  & $\langle 2f_{i+\epsilon}(x)\rangle$ \\
  $\langle 2 \rangle$   & $\langle 2\rangle$ \\
 $\langle 2f_i(x)\rangle$   & $\langle f_{i+\epsilon}(x)\rangle$ \\
  $\langle 0\rangle$   & $\langle 1\rangle$ \\
 $\langle f_i(x),v\rangle$   & $\langle 2vf_{i+\epsilon}(x)\rangle$ \\
 $\langle f_i(x)b(x)+v, 2\rangle$  & $\langle 2(-f_{i+\epsilon}(x)\delta_ix^{2n-m_i} b(x^{-1})+v)\rangle$ \\
 $\langle 2a(x)+v, 2f_i(x)\rangle$ & $\langle f_{i+\epsilon}(x)(2(1+a(x^{-1}))+v)\rangle$ \\
 $\langle 2, vf_i(x)\rangle$  & $\langle 2f_{i+\epsilon}(x), 2v\rangle$   \\
 $\langle 2f_i(x), 2v\rangle$   & $\langle 2, vf_{i+\epsilon}(x)\rangle$ \\
 $\langle 2b(x)+vf_i(x), 2f_i(x)\rangle$ & $\langle 2b(x^{-1})-vf_{i+\epsilon}(x)\delta_ix^{2n-m_i}, 2f_{i+\epsilon}(x)\rangle$ \\
 \hline
\end{tabular}
\end{center}

\noindent
\textit{where $a(x),b(x)\in \mathcal{T}_i$}.

\vskip 2mm \par
  (ii) \textit{If $1\leq i\leq \lambda$, $C_i$ is given by one of the following three cases}:

\vskip 2mm \par
  (ii-1) $C_i=\langle 2\rangle$.

\vskip 2mm \par
  (ii-2) \textit{$C_i=\langle 2(a(x)+b(x)f_i(x))+v\rangle$, where $(a(x),b(x))\in \mathcal{W}_i^{(1)}$}.

\vskip 2mm \par
  (ii-3) \textit{$C_i=\langle 2b(x)+vf_i(x), 2f_i(x)\rangle$, where $b(x)\in \mathcal{W}_i^{(2)}$}.

\vskip 2mm \par
  \textit{The number of self-dual negacyclic codes over $R$ of length $2n$ is equal to}
$$\prod_{i=1}^\lambda(1+|\mathcal{W}_i^{(1)}|+|\mathcal{W}_i^{(2)}|)
\prod_{j=1}^\epsilon(2^{2m_{\lambda+j}}+5\cdot 2^{m_{\lambda+j}}+9).$$

\vskip 3mm \noindent
  {\bf Proof.} By Lemma 5.2 (i) and (ii), we have that $\mu(i)=i$ for all $1\leq i\leq \lambda$,
$\mu(i)=i+\epsilon$ and $\mu(i+\epsilon)=i$ for all $\lambda+1\leq i\leq \lambda+\epsilon$. From this
by Theorem 5.5 and its proof, we deduce that
\begin{eqnarray*}
\mathcal{C}^{\bot}&=&(\bigoplus_{i=1}^\lambda\varepsilon_i(x)D_i)
 \oplus(\bigoplus_{i=\lambda+1}^{\lambda+\epsilon}(\varepsilon_{\mu(i)}(x)D_{\mu(i)}\oplus
 \varepsilon_{\mu(i+\epsilon)}(x)D_{\mu(i+\epsilon)}))\\
 &=&(\bigoplus_{i=1}^\lambda\varepsilon_i(x)D_i)
 \oplus(\bigoplus_{i=\lambda+1}^{\lambda+\epsilon}(\varepsilon_{i+\epsilon}(x)D_{i+\epsilon}\oplus
 \varepsilon_{i}(x)D_{i}))
\end{eqnarray*}
Then by Theorem 4.2, we see that $\mathcal{C}=\mathcal{C}^{\bot}$ if and only of
$C_i=D_i$ for all $i=1,\ldots,r$. Now, we have one of the following two cases.

\par
 \par
  (i) Let $i=\lambda+j$ where $1\leq j\leq \epsilon$. By $\mu(i)=i+\epsilon$,
we have $C_{i+\epsilon}=D_{i+\epsilon}=D_{\mu(i)}$. Then the conclusions follow from
Theorem 5.5 immediately.

\par
  (ii) $1\leq i\leq \lambda$. In this case, $\mu(i)=i$. Then by $|C_i|=|D_i|$ and Theorem 5.5, we have
one of the following three subcases:

\par
  (ii-1) $C_i=D_i=\langle 2\rangle$.

\par
  (ii-2) $C_i=\langle 2(a(x)+b(x)f_i(x))+v\rangle$ and $D_i=\langle v+2(1+a(x^{-1})-\delta_ix^{2n-m_i}b(x^{-1})f_i(x))\rangle$
as ideals of $\mathcal{K}_i+v\mathcal{K}_i$. Then by Theorem 4.4,
Theorem 2.5 and Theorem 3.2(iii), we see that $C_i=D_i$ if and only if
$a(x)=1+a(x^{-1})$
and $b(x)=-\delta_ix^{2n-m_i}b(x^{-1})$
as elements of the finite field $\mathbb{F}_2[x]/\langle \overline{f}_i(x)\rangle$. As
$-\delta_i=1$ in $\mathbb{F}_2$, these conditions are equivalent to
$(a(x),b(x))\in\mathcal{W}_i^{(1)}$.

\par
  (ii-3) $C_i=\langle 2b(x)+vf_i(x), 2f_i(x)\rangle$ and $D_i=\langle 2\widehat{b}(x)-vf_{i}(x)\delta_ix^{2n-m_i}$, $2f_{i}(x)\rangle$.
In this case, $C_i=D_i$ if and only if $b(x)=\frac{1}{-\delta_ix^{2n-m_i}}b(x^{-1})$ in  $\mathbb{F}_2[x]/\langle \overline{f}_i(x)\rangle$. Obviously, the latter is equivalent to
$b(x)\in\mathcal{W}_i^{(2)}$.
\hfill $\Box$



\section{Negacyclic codes over $\mathbb{Z}_{4}+v\mathbb{Z}_{4}$ of length $2M_p$}
  Recall that $2^p-1$ is called a \textit{Mersenne prime}, denoted by
  $$M_p=2^p-1,$$
if $p$ is a prime and $2^p-1$ is a prime as well. For example, when $p=2,3,5,7,13,17,19,31$, $M_p$ is a prime.
In this section, we present negacyclic codes over $\mathbb{Z}_{4}+v\mathbb{Z}_{4}$ ($v^2=2v$) of length $2M_p$.

\vskip 3mm \noindent
   {\bf Theorem 6.1} \textit{Let $n=M_p$ where $p\geq 3$. Then the number of all negacyclic
codes over $\mathbb{Z}_4+u\mathbb{Z}_4$ of length $2n$ and the number of self-dual negacyclic
codes over $\mathbb{Z}_4+u\mathbb{Z}_4$ of length $2n$ are equal to}
$$23\cdot(4^{p}+5\cdot 2^p+9)^{2\cdot\frac{2^{p-1}-1}{p}} \ {\rm and} \ 3\cdot(4^{p}+5\cdot 2^p+9)^{\frac{2^{p-1}-1}{p}},
\ {\rm respectively}.$$

 \noindent
   {\bf Proof.} Since $n=M_p$ and $M_p$ is a Mersenne prime, for any integer $s$, $1\leq s\leq n-1$, the $2$-cyclotomic coset modulo $n$
containing $s$ is $J^{(2)}_s=\{s,2s,\ldots,2^{p-1}s\}$ with $|J^{(2)}_s|=p$. Therefore, the number of $2$-cyclotomic cosets modulo $n$
is equal to $1+\frac{n-1}{p}=1+\frac{2^p-1-1}{p}=1+2\cdot\frac{2^{p-1}-1}{p}$.

Suppose that $2^js\equiv -s$ (mod $n$) for some
$1\leq j\leq p-1$. Then $(2^p-1)|s(2^j+1)$. Since ${\rm gcd}(s,2^p-1)=1$, we have $(2^p-1)|(2^j+1)$. This implies $2^p-1\leq 2^j+1$. On the other hand,
by $p\geq 3$ and $1\leq j<p$, we have $(2^p-1)-(2^j+1)=2^j(2^{p-j}-1)-2>0$. This implies $(2^p-1)>(2^j+1)$, and we get a contradiction. Therefore,
$J^{(2)}_{-s}\neq J^{(2)}_s$ for all $1\leq s\leq n-1$.

\par
   Using the notations of Lemma 5.2(ii), we have
$r=1+2\cdot\frac{2^{p-1}-1}{p}$, $\lambda=1$, $\epsilon=\frac{2^{p-1}-1}{p}$, $m_1=|J^{(2)}_0|=1$
and $m_i=|J^{(2)}_1|=p$ for all $i=2,3,\ldots,r$.

\par
  By Corollary 4.5, the number of all negacyclic
codes over $\mathbb{Z}_4+v\mathbb{Z}_4$ of length $2n$ is 
$(2^2+5\cdot 2+9)\prod_{i=2}^r(2^{2p}+5\cdot 2^p+9)=23\cdot(4^{p}+5\cdot 2^p+9)^{2\cdot\frac{2^{p-1}-1}{p}}.$

\par
   It is obvious that $f_1(x)=x-1$ and
$\widetilde{f}_1(x)=1-x=\delta_1f_1(x)$ with $\delta_1=-1$.
Hence ${\cal K}_1=\mathbb{Z}_4[x]/\langle f_1(-x^2)\rangle=\mathbb{Z}_4[x]/\langle x^2+1\rangle$. Using the notations
in Section 3, by $m_1=1$ we have ${\cal T}_1=\{0,1\}$ and $\mathbb{F}_2[x]/\langle \overline{f}_1(x)\rangle=\mathbb{F}_2$ where $\overline{f}_1(x)=x-1\in \mathbb{F}_2[x]$. Then we have

\par
  $\mathcal{W}_1^{(1)}=\{(a,b)\mid a+a+1=b+b=0, \ a,b\in \mathbb{F}_2\}=\emptyset$ with $|\mathcal{W}_1^{(1)}|=0$;

\par
  $\mathcal{W}_1^{(2)}=\{b\in \mathbb{F}_2\mid b+b=0\}=\mathbb{F}_2$ with $|\mathcal{W}_1^{(2)}|=2$.

\noindent
  Therefore, using the notations in the proof of Theorem 4.6(i) we see that
there are $3$ ideals of $\mathcal{K}_1+v\mathcal{K}_1$ satisfying $C_1=D_1$:
  $$\langle 2\rangle, \ \langle v(x-1), 2(x-1)\rangle, \ \langle 2+v(x-1), 2(x-1)\rangle.$$

\par
  For any $i=1+j$, where $1\leq j\leq \epsilon=\frac{2^{p-1}-1}{p}$, By Theorem 4.4
and $m_i=p$ we know that the number of pairs of $(C_i,C_{i+\epsilon})$ in Theorem 5.6(ii) is equal to
$2^{2p}+5\cdot 2^p+9=4^{p}+5\cdot 2^p+9$.

\par
  As stated above, by Theorem 5.6 we conclude that the number of self-dual negacyclic codes
over $\mathbb{Z}_4+v\mathbb{Z}_4$ of length $2(2^p-1)$, where both $p$ and $2^p-1$ are prime integers,
is equal to $3\cdot(4^{p}+5\cdot 2^p+9)^{\frac{2^{p-1}-1}{p}}$.
\hfill $\Box$

\vskip 3mm\par
   For example, when $p=5$, we have $M_5=31$. Hence the number of
 self-dual negacyclic codes over $\mathbb{Z}_4+v\mathbb{Z}_4$
of length $62$ is equal to
$$3\cdot (4^{5}+5\cdot 2^5+9)^{\frac{2^{5-1}-1}{5}}
=3\cdot 1193^3=5093808171\approx 5\times 10^9.$$
All these codes can be listed by Theorem 5.6. From these codes, by Proposition 1.3 we obtain
$5093808171$ self-dual $2$-quasi-twisted codes over $\mathbb{Z}_4$
of length $124$, and $5093808171$ self-dual $4$-quasi-twisted binary codes
of length $248$ by the Gray map from $\mathbb{Z}_4$ onto $\mathbb{F}_2^2$: $0\mapsto 00, 1\mapsto 01, 2\mapsto 11, 3\mapsto 10$.

\par
   When $p=7$,  we have $M_7=127$. Hence the number of
 self-dual negacyclic codes over $\mathbb{Z}_4+v\mathbb{Z}_4$
of length $254$ is equal to
$$
3\cdot (4^{7}+5\cdot 2^7+9)^{\frac{2^{7-1}-1}{7}}
=3\cdot 17033^9\approx 3.62\times 10^{38}.$$
All these codes can be listed by Theorem 5.6.

\par
  Now, let $p=3$. Then $n=M_3=2^3-1=7$. By Theorem 6.1
we know that the number of all negacyclic
codes over $\mathbb{Z}_4+v\mathbb{Z}_4$ of length $14$ is
\begin{center}
$23\cdot (4^{3}+5\cdot 2^3+9)^2=23\cdot 113^2=293687$,
\end{center}
and the number of self-dual negacyclic
codes over $\mathbb{Z}_4+v\mathbb{Z}_4$ of length $14$ is equal to
$3\cdot (4^{3}+5\cdot 2^3+9)=3\cdot 113=339$.

\par
   Precisely, we have that
$x^7-1=f_1(x)f_2(x)f_3(x),$
where $f_1(x)=x-1$, $f_2(x)=x^3+2x^2+x+3$ and $f_3(x)=x^3+3x^2+2x+3$ are
pairwise coprime basic irreducible polynomials in $\mathbb{Z}_{4}[x]$. Hence
$r=3$, $m_1=1$ and $m_2=m_3=3$. Obviously, $\widetilde{f}_1(x)=\delta_1f_1(x)$ and $\widetilde{f}_2(x)=\delta_2f_3(x)$ where $\delta_1=\delta_2=-1$, which implies that
$\mu(1)=1$ and $\mu(2)=3$.

\par
  For each integer $i$, $1\leq i\leq 3$, we denote
$F_i(x)=\frac{x^7-1}{f_i(x)}$, and find polynomials $a_i(x),b_i(x)\in \mathbb{Z}_{4}[x]$
satisfying $a_i(x)F_i(x)+b_i(x)f_i(x)=1$. Then we set $\varepsilon_i(x)\equiv a_i(-x^2)F_i(-x^2)$ (mod $x^{14}+1$).
Precisely, we have

\par
   $\varepsilon_1(x)=3+x^2+3x^4+x^6+3x^8+x^{10}+3x^{12}$;

\par
   $\varepsilon_2(x)=1+x^2+3x^4+2x^6+3x^8+2x^{10}+2x^{12}$;

\par
  $\varepsilon_3(x)=1+2x^2+2x^4+x^6+2x^8+x^{10}+3x^{12}$.

\noindent
  Using the notations in Section 3,  we have ${\cal K}_i=\mathbb{Z}_4[x]/\langle f_i(-x^2)\rangle$
for $i=1,2,3$, and
${\cal T}_2=\{c_0+c_1x+c_2x^2\mid c_0,c_1,c_2\in \{0,1\}\}$. As $x^{14}=-1$,
we have $x^{-1}=-x^{13}$ in ${\cal K}_2$.
Hence for any $a(x)=a_0+a_1x+a_2x^2\in {\cal T}_2$, it follows that
$$
a(x^{-1})
 =a_0+2(a_1+a_2)+3a_2x^2+3a_1x^3+3a_2x^4+3a_1x^5
 \ ({\rm mod} \ f_3(-x^2))
$$
and $\overline{a}(x^{-1})=a_0+a_2x^2+a_1x^3+a_2x^4+a_1x^5$.
By Theorem 5.6, all distinct self-dual negacyclic codes over $\mathbb{Z}_{4}+v\mathbb{Z}_{4}$ of length $14$
are given by
$${\cal C}=\varepsilon_1(x)C_1\oplus \varepsilon_2(x)C_2\oplus \varepsilon_3(x)C_3 \ ({\rm mod} \ x^{14}+1),$$
where $C_i$ is an ideal of the ring ${\cal K}_i+v{\cal K}_i$
(${\cal K}_i=\mathbb{Z}_4[x]/\langle f_i(-x^2)\rangle$) satisfying one of the following conditions:

\vskip 2mm \par
  $\bullet$ $C_1$ is one of the following $3$ ideals: $\langle 2\rangle$,
$\langle v(x-1), 2(x-1)\rangle$, $\langle 2+v(x-1), 2(x-1)\rangle$.

\vskip 2mm \par
  $\bullet$ $(C_2,C_3)$ is given by one of the following $113$ cases, where $a(x),b(x)\in \mathcal{T}_2$:
\begin{center}
\begin{tabular}{ll}\hline
  $C_2$ (mod $f_2(-x^2)$) &  $C_{3}$ (mod $f_{3}(-x^2)$) \\ \hline
 $\langle 2(a(x)+b(x)f_2(x))+v\rangle$ &  $\langle 2(1+a(x^{-1})+x^{11}b(x^{-1})f_{3}(x))
 +v\rangle$  \\
  $\langle 2vf_2(x)\rangle$ & $\langle f_{3}(x),v\rangle$ \\
  $\langle 2(f_2(x) b(x)+v)\rangle$ & $\langle f_{3}(x)x^{11}b(x^{-1})+v, 2\rangle$ \\
 $\langle f_2(x)(2a(x)+v)\rangle$   & $\langle 2(1+a(x^{-1}))+v, 2f_{3}(x)\rangle$ \\
  $\langle 1\rangle$   & $\langle 0\rangle$ \\
  $\langle f_2(x)\rangle$  & $\langle 2f_{3}(x)\rangle$ \\
  $\langle 2 \rangle$   & $\langle 2\rangle$ \\
 $\langle 2f_2(x)\rangle$   & $\langle f_{3}(x)\rangle$ \\
  $\langle 0\rangle$   & $\langle 1\rangle$ \\
 $\langle f_2(x),v\rangle$   & $\langle 2vf_{3}(x)\rangle$ \\
 $\langle f_2(x)b(x)+v, 2\rangle$  & $\langle 2(f_{3}(x)x^{11} b(x^{-1})+v)\rangle$ \\
 $\langle 2a(x)+v, 2f_2(x)\rangle$ & $\langle f_{3}(x)(2(1+a(x^{-1}))+v)\rangle$ \\
 $\langle 2, vf_2(x)\rangle$  & $\langle 2f_{3}(x), 2v\rangle$   \\
 $\langle 2f_2(x), 2v\rangle$   & $\langle 2, vf_{3}(x)\rangle$ \\
 $\langle 2b(x)+vf_2(x), 2f_2(x)\rangle$ & $\langle 2b(x^{-1})+vf_{3}(x)x^{11}, 2f_{3}(x)\rangle$ \\
 \hline
\end{tabular}
\end{center}

\par
 Among the $339$ self-dual negacyclic codes over $\mathbb{Z}_{4}+v\mathbb{Z}_{4}$ of length $14$, we obtain $36$ codes $\mathcal{C}$ whose Gray image $\theta(\mathcal{C})$ are new good self-dual $2$-quasi-twisted code over $\mathbb{Z}_4$ of length $28$ (these codes does not exist in [21] too). The new codes we obtained have new type, better minimum Lee weight, more algebraic structures and properties.

\par
  $\diamondsuit$ When $C_1=\langle v(x-1), 2(x-1)\rangle$ or $C_1=\langle 2+v(x-1), 2(x-1)\rangle$, and $(C_2,C_3)$ is given by one of the following $12$ cases:
\begin{center}
\begin{tabular}{ll}\hline
  $C_2$ (mod $f_2(-x^2)$) &  $C_{3}$ (mod $f_{3}(-x^2)$) \\ \hline
  $\langle 2(x^2f_2(x)+v)\rangle$ & $\langle f_{3}(x)x^{11}(x^4 + x^2)+v, 2\rangle$ \\
   $\langle 2(xf_2(x)+v)\rangle$ & $\langle f_{3}(x)x^{11}(x^5 + x^3)+v, 2\rangle$ \\
    $\langle 2((x^2+x)f_2(x)+v)\rangle$ & $\langle f_{3}(x)x^{11}(x^5 + x^4 + x^3 + x^2)+v, 2\rangle$ \\
     $\langle 2((x^2+1)f_2(x)+v)\rangle$ & $\langle f_{3}(x)x^{11}(x^4 + x^2 + 1)+v, 2\rangle$ \\
      $\langle 2((x+1)f_2(x)+v)\rangle$ & $\langle f_{3}(x)x^{11}(x^5 + x^3 + 1)+v, 2\rangle$ \\
       $\langle 2((x^2+x+1)f_2(x)+v)\rangle$ & $\langle f_{3}(x)x^{11}(x^5 + x^4 + x^3 + x^2 + 1)+v, 2\rangle$ \\
    $\langle x^2f_2(x)+v, 2\rangle$  & $\langle 2(f_{3}(x)x^{11}(x^4 + x^2 + 1)+v)\rangle$ \\
  $\langle xf_2(x)+v, 2\rangle$  & $\langle 2(f_{3}(x)x^{11}(x^5 + x^3)+v)\rangle$ \\
   $\langle (x^2+x)f_2(x)+v, 2\rangle$  & $\langle 2(f_{3}(x)x^{11}(x^5 + x^4 + x^3 + x^2)+v)\rangle$ \\
    $\langle (x^2+1)f_2(x)+v, 2\rangle$  & $\langle 2(f_{3}(x)x^{11}(x^4 + x^2 + 1)+v)\rangle$ \\
     $\langle (x+1)f_2(x)+v, 2\rangle$  & $\langle 2(f_{3}(x)x^{11}(x^5 + x^3 + 1)+v)\rangle$ \\
      $\langle (x^2+x+1)f_2(x)+v, 2\rangle$  & $\langle 2(f_{3}(x)x^{11}(x^5 + x^4 + x^3 + x^2 + 1)+v)\rangle$ \\
 \hline
\end{tabular}
\end{center}
we obtain $24$ new good self-dual $\mathbb{Z}_{4}$ codes
$\theta(\mathcal{C})$ with basic parameters $(28, 2^{28}$, $d_L=8, d_E=12)$ and of type $2^{14}4^7$, where $d_L$ is the minimum Lee weight and $d_E$ is the minimum Euclidean weight of $\mathcal{C}$.

\par
  $\diamondsuit$ When $C_1=\langle 2\rangle$ and $(C_2,C_3)$ is given by one of the above $12$ cases,
we obtain $12$ new good self-dual $\mathbb{Z}_{4}$ codes
$\theta(\mathcal{C})$ with basic parameters $(28, 2^{28}$, $d_L=6, d_E=12)$ and of type $2^{16}4^6$.

\vskip 3mm\noindent
  {\bf Remark}  The existing self-dual codes over $\mathbb{Z}_4$ are listed in the database [26] and the maximal code length is $19$ in the table. In [27], the existing linear codes over $\mathbb{Z}_{4}$ of length $24$ have basic parameters $(24, 2^{28}, d_L=6)$ and $(24, 2^{28}, d_L=5)$ and both of type $4^{14}$.


\section{Conclusions and further research} \label{}
\noindent
We have developed a theory for negacyclic codes over the ring $\mathbb{Z}_4+v\mathbb{Z}_4$ ($v^2=2v$) of oddly even length,
including the enumeration and construction of
these codes, the dual code and self-duality for each of these codes. These codes enjoy a rich algebraic structure 
(which makes the search process much simpler) compared
to arbitrary linear codes. Our further interest is to consider
negacyclic codes over $\mathbb{Z}_4+v\mathbb{Z}_4$ of arbitrary even length.
On the other hand, obtaining some bounds for minimum distance such as BCH-like of a negacyclic code over the ring $\mathbb{Z}_4+v\mathbb{Z}_4$ by just looking at the canonical form decomposition would
be rather interesting.

\vskip 3mm \noindent {\bf Acknowledgments}
Part of this work was done when Yonglin Cao was visiting Chern Institute of Mathematics, Nankai University, Tianjin, China. Yonglin Cao would like to thank the institution for the kind hospitality. This research is
supported in part by the National Natural Science Foundation of
China (Grant Nos. 11671235, 11471255).

\vskip 5mm \noindent
{\bf Appendix: Proof of Theorem 2.5}

\vskip 3mm
   Using the notations of Section 2, by [6] Example 2.5 we know that the number of
linear codes over the finite chain ring $\mathcal{K}$ of length $2$ is equal to
$\sum_{i=0}^4(2i+1)|\mathcal{T}|^{4-i}$.
Moreover, every linear code $C$ over
$\mathcal{K}$ of length $2$ has one and only one of the following matrices $G$ as their generator matrices:

\vskip 2mm \par
 (i) \textit{$G=(1,a)$, $a\in \mathcal{K}$}.
\ \ \ \ \ \ (ii) \textit{$G=(\pi^k,\pi^ka)$, $a\in \mathcal{K}/\langle \pi^{4-k}\rangle$, $1\leq k\leq 3$}.

 \par
(iii) \textit{$G=(\pi b,1)$, $b\in \mathcal{K}/\langle \pi^{3}\rangle$}.

 \par
(iv) \textit{$G=(\pi^{k+1}b,\pi^k)$, $b\in \mathcal{K}/\langle \pi^{4-k-1}\rangle$, $1\leq k\leq 3$}.

 \par
 (v) \textit{$G=\pi^kI_2$, $0\leq k\leq 4$}.
\ \
  (vi) \textit{$G=\left(\begin{array}{cc}1 & c\cr
0 & \pi^t\end{array}\right)$,  $c\in \mathcal{K}/\langle \pi^{t}\rangle$, $1\leq t\leq 3$}.

 \par
  (vii) \textit{$G=\left(\begin{array}{cc} \pi^k & \pi^kc\cr
0 & \pi^{k+t}\end{array}\right)$,  $c\in \mathcal{K}/\langle \pi^{t}\rangle$, $1\leq t\leq 4-k-1$, $1\leq k\leq 2$}.

 \par
    (viii) \textit{$G=\left(\begin{array}{cc}c & 1\cr \pi^t & 0\end{array}\right)$, where $c\in \pi(\mathcal{K}/\langle \pi^{t}\rangle)$ and $1\leq t\leq 3$}.

 \par
    (ix) \textit{$G=\left(\begin{array}{cc}\pi^kc & \pi^k\cr \pi^{k+t} & 0\end{array}\right)$, $c\in \pi(\mathcal{K}/\langle \pi^{t}\rangle)$,
$1\leq t\leq 4-k-1$, $1\leq k\leq 2$}.

\vskip 3mm \noindent
Therefore, we only need to consider the nine cases listed above:

\vskip 2mm\par
   (i) Suppose that $C$ satisfies Condition (2). By $(1,a)\in C$, we have
$(0,1+\omega \pi^2a)\in C$. Since $G$ is the generator matrix of $C$, there exists $b\in \mathcal{K}$ such that $(0,1+\omega \pi^2a)=b(1,a)=(b,ba)$,
i.e. $0=b$ and $1+\omega \pi^2a=ba$. This implies $1+\omega \pi^2a=0$. So we get a contradiction, since $1+\omega \pi^2a$ is an invertible element
of $\mathcal{K}$. Hence $C$ does not satisfy Condition (2) in this case.

\par
   (ii) Suppose that $C$ satisfies Condition (2). By $(\pi^k,\pi^ka)\in C$,
we have $(0,\pi^k+\omega\pi^{k+2})=(0,\pi^k+\omega \pi^2\cdot\pi^{k}a)\in C$. Then there exists $b\in \mathcal{K}$ such that $
(0,\pi^k+\omega \pi^{k+2}a)=b(\pi^k,\pi^k a)=(\pi^k b,\pi^k ba)$.
 This implies $0=\pi^k b$ and $\pi^k(1+\omega\pi^2a) =\pi^k ba$. Hence $\pi^k(1+\omega\pi^2a) =0$. Since $1+\omega\pi^2a$ is an invertible element
of $\mathcal{K}$, we deduce $\pi^k=0$, which contradict that $k\leq 3$.

\par
   (iii) In this case, $C$ satisfies Condition (2) if and only if
there exists $a\in \mathcal{K}$ such that $(0,\pi b+\omega\pi^2)=(0,\pi b+\omega\pi^2\cdot 1)=a(\pi b,1)=(\pi ab,a)$, i.e. $0=\pi ab$ and $\pi(b+\omega\pi)=\pi b+\omega\pi^2=a$. These conditions
are equivalent to that $b$ satisfies $\pi^2(b+\omega\pi)b=0$, i.e. $(b+\omega\pi)b\in \pi^{2}\mathcal{K}$.
The latter condition is equivalent to $\|(b+\omega\pi)b\|_\pi\geq 2$. As $b\in \mathcal{K}/\langle \pi^3\rangle$, $b$ has a unique $\pi$-expansion:
$b=r_0+r_1\pi+r_2\pi^2$ with $r_0,r_1,r_{2}\in \mathcal{T}$. This implies $b+\omega\pi=r_0+(\omega+r_1)\pi+r_2\pi^2$.
Suppose that $r_0\neq 0$. Then $\|b\|_\pi=\|b+\omega\pi\|_\pi=0$. This implies $\|(b+\omega\pi)b\|_\pi=0$ and we get a contradiction. Now, let $r_0=0$. Then by $(b+\omega\pi)b=\pi^2(\omega+r_1+r_2\pi)(r_1+r_2\pi)$
it follow that $\|(b+\omega\pi)b\|_\pi\geq 2$. Hence $b=\pi(r_1+r_2\pi)$ for all $r_1,r_{2}\in \mathcal{T}$
in this case.

\par
   (iv) In this case, $C$ satisfies Condition (2) if and only if
there exists $a\in \mathcal{K}$ such that $(0,\pi^{k+1}(b+\omega\pi))=(0,\pi^{k+1}b+\omega\pi^2\cdot \pi^{k})=a(\pi^{k+1}b,\pi^k)=(\pi^{k+1}ab,\pi^ka)$, i.e. $0=\pi^{k+1}ab=\pi b\cdot\pi^ka$ and $\pi^{k+1}(b+\omega\pi)=\pi^ka$. These conditions
are equivalent to that $b$ satisfies $\pi^{k+2}(b+\omega\pi)b=0$, i.e.
$\|(b+\omega\pi)b\|_\pi\geq 4-k-2$
where $1\leq k\leq 3$. Then we have one of the following three cases:

\par
  (iv-1) Let $k=3$. Then $\|(b+\omega\pi)b\|_\pi\geq 4-k-2=-1$ for any $b\in \mathcal{K}/\langle \pi^{4-3-1}\rangle=\mathcal{K}/\langle 1\rangle=\{0\}$. Hence $G=(0,\pi^{3})$.

\par
  (iv-2) Let $k=2$. $\|(b+\omega\pi)b\|_\pi\geq 4-k-2=0$ for any $b\in \mathcal{K}/\langle \pi^{4-2-1}\rangle=\mathcal{K}/\langle \pi\rangle=\mathcal{T}$. In this case, we have
$G=(\pi^{3}b,\pi^{2})$.

\par
  (iv-3) Let $k=1$. $\|(b+\omega\pi)b\|_\pi\geq 4-k-2=1$ if and only if $b\in \pi(\mathcal{K}/\langle \pi^{4-1-1}\rangle)=\pi(\mathcal{K}/\langle \pi^{2}\rangle)$. In this case, we have
$G=(\pi^2b,\pi)=(\pi^{3}a,\pi)$ for all $b=\pi a$ with $a\in \mathcal{T}$.

\par
   (v) In this case, $C$ satisfies Condition (2) for all $0\leq k\leq 4$.

\par
   (vi) Suppose
that $C$ satisfies Condition (2). Then there exist $a,b\in \mathcal{K}$ such that $(0,1+\omega\pi^2c)=a(1,c)+b(0,\pi^t)=(a,ac+\pi^t b)$,
i.e. $0=a$ and $1+\omega\pi^2c=ac+\pi^t b$. This implies $1=\pi(\pi^{t-1} b-\omega\pi c)$ and we get a contradiction.
Hence $C$ does not satisfy Condition (2) in this case.

\par
   (vii) Suppose
that $C$ satisfies Condition (2). Then there exist $a,b\in \mathcal{K}$ such that $(0,\pi^k+\omega\pi^2\cdot \pi^kc)=a(\pi^k,\pi^kc)+b(0,\pi^{k+t})
=(\pi^ka,\pi^kac+\pi^{k+t}b)$,
i.e. $0=\pi^ka$ and $\pi^k+\omega\pi^{k+2}c=\pi^ka\cdot c+\pi^{k+t} b$. This implies $\pi^k=\pi^{k+1}(\pi^{t-1}b-c\omega\pi)$, and we get a contradiction
as $1\leq k\leq 3$.

\par
   (viii) It is clear that $(0,\pi^t)=\pi^t(c,1)-c(\pi^t,0)\in C$.
Hence $C$ satisfies Condition (2) if and only if there exist $a,b\in \mathcal{K}$ such that
$(0,c+\omega\pi^2\cdot 1)=a(c,1)+b(\pi^t,0)=(ac+\pi^tb,a)$,
i.e. $0=ac+\pi^tb$ and $c+\omega\pi^2=a$, which are equivalent to
that $(c+\omega\pi^2)c=-\pi^tb\in \pi^t\mathcal{K}$ for some $b\in \mathcal{K}$, i.e. $\|(c+\omega\pi^2)c\|_\pi\geq t$, where $c\in \pi(\mathcal{K}/\langle \pi^t\rangle)$ and $1\leq t\leq 3$.
Then we have one of the following three subcases:

\par
   (viii-1) When $t=1$, then $c\in \pi(\mathcal{K}/\langle \pi\rangle)=\{0\}$, i.e. $c=0$. In this case, $\|(c+\omega\pi^2)c\|_\pi=\|0\|_\pi=4>t$. Hence $G=\left(\begin{array}{cc}0 & 1\cr \pi & 0\end{array}\right)$.

\par
   (viii-2) When $t=2$, then $c\in \pi(\mathcal{K}/\langle \pi^2\rangle)=\pi\mathcal{T}$, i.e. $c=\pi z$
where $z\in \mathcal{T}$. In this case, $\|(c+\omega\pi^2)c\|_\pi=\|(\pi h+\omega\pi^2)\cdot \pi h\|_\pi\geq 2=t$. Hence $G=\left(\begin{array}{cc}\pi z & 1\cr \pi^2 & 0\end{array}\right)$.

\par
   (viii-3) When $t=3$, then $c\in \pi(\mathcal{K}/\langle \pi^3\rangle)$, i.e. $c=r_1\pi+r_2\pi^2$
where $r_1,r_2\in \mathcal{T}$. Suppose $r_1\neq 0$. Then $\|c\|_\pi=\|c+\omega\pi^2\|_\pi=1$.
This implies $\|(c+\omega\pi^2)c\|_\pi$ $=2<3=t$. Now, let $r_1=0$. Then
it is obvious that $\|(c+\omega\pi^2)c\|_\pi=4>t$ for all $c=\pi^2z$ where $z=r_2\in \mathcal{T}$.
 Hence $G=\left(\begin{array}{cc}\pi^2 z & 1\cr \pi^3 & 0\end{array}\right)$.

\par
   (xi) It is clear that $(0,\pi^{k+t})=\pi^t(\pi^kc,\pi^k)-c(\pi^{k+t},0)\in C$.
Hence $C$ satisfies Condition (2) if and only if there exist $a,b\in \mathcal{K}$ such that
$(0,\pi^kc+\omega\pi^2\cdot \pi^k)=a(\pi^kc,\pi^k)+b(\pi^{k+t},0)=(\pi^kac+\pi^{k+t}b,\pi^ka)$,
i.e. $0=\pi^kac+\pi^{k+t}b$ and $\pi^k(c+\omega\pi^2)=\pi^ka$, which are equivalent to
that $\pi^k(c+\omega\pi^2)c=-\pi^{k+t}b\in \pi^{k+t}\mathcal{K}$, i.e. $\|(c+\omega\pi^2)c\|_\pi\geq t$, where $c\in \pi(\mathcal{K}/\langle \pi^t\rangle)$, $1\leq t\leq 3-k$ and $1\leq k\leq 2$.
Then we have one of the following two subcases:

\par
  (xi-1) Let $k=1$. Then $1\leq t\leq 2$. If $t=1$, then $c\in \pi(\mathcal{K}/\langle \pi\rangle)=\{0\}$, i.e. $c=0$. Obviously, $\|(c+\omega\pi^2)c\|_\pi=4>t$. Hence
$G=\left(\begin{array}{cc}0 & \pi\cr \pi^{2} & 0\end{array}\right)$.

\par
  If $t=2$, then $c\in \pi(\mathcal{K}/\langle \pi^2\rangle)=\pi \mathcal{T}$, i.e. $c=\pi z$
where $z\in \mathcal{T}$. Obviously, $\|(c+\omega\pi^2)c\|_\pi=2\geq t$. Hence
$G=\left(\begin{array}{cc} \pi^2z & \pi\cr \pi^{3} & 0\end{array}\right)$.

\par
  (xi-2) Let $k=2$. Then $t=1$ and $c=0$. Hence $G=\left(\begin{array}{cc}0 & \pi^2\cr \pi^{3} & 0\end{array}\right)$.


\begin{thebibliography}{s20}
%
%

\bibitem{s1} Abualrub T., Siap I.: Cyclic codes over the ring $\mathbb{Z}_2+u\mathbb{Z}_2$ and $\mathbb{Z}_2+u\mathbb{Z}_2+u^2\mathbb{Z}_2$,
Des. Codes Cryptogr. {\bf 42} (2007), 273--287.

\bibitem{s2} Bonnecaze, A., Udaya, P: Cyclic codes and self-dual codes
over $\mathbb{F}_2+u\mathbb{F}_2$, IEEE Trans. Inform. Theory {\bf
45} (1999), 1250--1255.

\bibitem{s3} Calderbank, A.R., Hammons Jr., A.R., Kumar, P.V., Sloane, N.J.A. and Sol\'{e}, P.:  A linear
construction for certain Kerdock and Preparata codes, Bulletin of the American Mathematical
Society {\bf 29}(2) (1993), 218--222.

\bibitem{s4} Calderbank, A.R., Hammons Jr., A.R., Kumar, P.V., Sloane, N.J.A. and Sol\'{e}, P.:  The
$Z_4$-linearity of Kerdock, Preparata, Goethals, and related codes, IEEE Trans. Inform. Theory {\bf 40} (1994), 301--319.

\bibitem{s5} Cao, Y., Cao, Y.: Complete classification for simple root cyclic codes over local rings $\mathbb{Z}_{p^s}[v]/\langle v^2-pv\rangle$,\\
     https://www.researchgate.net/publication/320620031.

\bibitem{s6} Cao, Y., Gao, Y.: Repeate root cyclic $\mathbb{F}_q$-linear codes over
$\mathbb{F}_{q^l}$,
Finite Fields Appl. {\bf 31} (2015), 202--227.

\bibitem{s7} Cao, Y.:  On constacyclic codes over finite chain rings, Finite
Fields Appl. {\bf 24} (2013), 124--135.

\bibitem{s8} Cao, Y., Cao, Y., Li, Q.: The concatenated structure of cyclic codes over
$\mathbb{Z}_{p^2}$, J. Appl. Math. Comput. {\bf 52} (2016), 363--385.

\bibitem{s9} Cao, Y., Cao, Y., Li, Q.: Concatenated structure of cyclic codes over
$\mathbb{Z}_4$ of length $4n$, Appl. Algebra in Engrg. Comm. Comput.
{\bf 10} (2016), 279--302.

\bibitem{10} Dinh, H. Q., L\'{o}pez-Permouth, S. R.:
Cyclic and negacyclic codes over finite chain rings, IEEE Trans.
Inform. Theory {\bf 50} (2004), 1728--1744.

\bibitem{s11} Dinh, H. Q.: Negacyclic codes of length $2^s$ over
Galois rings, IEEE Trans. Inform. Theory {\bf 51} (2005),
4252--4262.

\bibitem{s12} Dinh H. Q.: Complete distance of all negacyclic codes of length $2^s$
over $\mathbb{Z}_{2^a}$. IEEE Trans. Inf. Theory {\bf 53} (2007), 147--161.

\bibitem{s13} Dougherty S. T., Kim J-L, Kulosman H., Liu H.: Self-dual
codes over commutative Frobenius rings, Finite Fields Appl. {\bf 16}, 14--26 (2010).

\bibitem{s14} Kanwar, P. and L\'{o}pez-Permouth, S. R.:  Cyclic codes over the integers modulo $p^m$,
Finite Fields Appl. {\bf 4}:3 (1997), 334--352.

\bibitem{s15} Mart\'{\i}nez-Moro, E., Szabo, S., Yildiz, B.: Linear codes over $\frac{\mathbb{Z}_4[x]}{\langle x^2+2x\rangle}$, Int. J. Information and Coding Theory, Vol. 3, No. 1 (2015), 78--96.

\bibitem{s16} Norton, G.,  S\u{a}l\u{a}gean-Mandache, A.: On the structure of linear and cyclic
codes over finite chain rings, Appl. Algebra in Engrg. Comm. Comput.
{\bf 10} (2000), 489--506.

\bibitem{s17} Shi, M., Zhang, Y.: Quasi-twisted codes with constacyclic constituent codes,
Finite Fields Appl. {\bf 39} (2016), 159--178.

\bibitem{s18} Shi, M., Guan, Y., Sol\'{e}, P.: Two new families of two-weight codes,
IEEE Trans.
Inform. Theory {\bf 63}(10) (2017), 6240--6246.

\bibitem{s18} Shi, M., Xu, L., Yang, G.: A note on one weight and two weight projec-
tive $\mathbb{Z}_4$-codes, IEEE Trans.
Inform. Theory {\bf 63}(1) (2017), 177--182.

\bibitem{s20} Shi, M., Sol\'{e}, P., Wu, B.: Cyclic codes and the weight enumerators over
$\mathbb{F}_2 +v\mathbb{F}_2 +v^2\mathbb{F}_2$, Applied and Computational Mathematics {\bf 12}(2) (2013),
247--255.

\bibitem{s21} Shi, M., Qian, L., Sok, L.,  Aydin, N.,  Sol\'{e}, P.:
On constacyclic codes over $\mathbb{Z}_4[u]/\langle u^2-1\rangle$ and
their Gray images, Finite Fields Appl. {\bf 45}:3 (2017), 86--95.

\bibitem{s22} Wan Z.-X.:  Quaternary Codes. World Scientific Pub Co Inc. 1997.

\bibitem{s23} Wood, J. A.: Duality for modules over finite rings and applications to coding theory,
American Journal of Mathematics, Vol. 121, No. 3 (1999), 555--575.

\bibitem{s24} Yildiz B., Karadeniz S.: Linear codes over $\mathbb{Z}_4+u\mathbb{Z}_4$: MacWilliams identities, projections, and formally self-dual codes,
Finite Fields Appl. {\bf 27}, 24--40 (2014).

\bibitem{s25} Yildiz, B. and Aydin, N.:  Cyclic codes over $\mathbb{Z}_4+u\mathbb{Z}_4$ and
$\mathbb{Z}_4$ images, International Journal
of Information and Coding Theory, Vol. 2, No. 4 (2014), 226--237.

    \bibitem{s26} Self-dual codes over integers modulo 4 (accessed on 04 Dec. 2017), \\ \url{http://www.math.is.tohoku.ac.jp/~Emunemasa/research/codes/sd4.htm}.

\bibitem{s27} The $\mathbb{Z}_{4}$ database (accessed on 04 Dec. 2017), \url{http://www.asamov.com/Z4Codes/CODES/ShowCODESTablePage.aspx}.

\end{thebibliography}


\end{document}